\PassOptionsToPackage{unicode}{hyperref}
\PassOptionsToPackage{hyphens}{url}
\PassOptionsToPackage{dvipsnames,svgnames,x11names}{xcolor}

\documentclass[aoas,preprint]{imsart}
\RequirePackage{amsthm,amsmath,amsfonts,amssymb}
\RequirePackage[authoryear]{natbib}
\RequirePackage[colorlinks,citecolor=blue,urlcolor=blue,linkcolor=blue]{hyperref}
\RequirePackage{graphicx}
\RequirePackage{url}

\RequirePackage{booktabs}
\RequirePackage{multirow}
\RequirePackage{float}

\RequirePackage{microtype}
\RequirePackage{xcolor}

\definecolor{TabGreen}{HTML}{2CA02C}
\definecolor{TabBlue}{HTML}{1F77B4}
\definecolor{TabOrange}{RGB}{242,142,43}

\RequirePackage{tikz}
\RequirePackage{tikz-3dplot}
\usetikzlibrary{decorations.pathreplacing}
\usetikzlibrary{calc}
\usetikzlibrary{matrix, positioning}

\RequirePackage{algorithm}
\RequirePackage[noend]{algpseudocode}
\algtext*{EndIf}
\algtext*{EndFor}
\algtext*{EndWhile}
\algtext*{EndProcedure}
\algtext*{EndFunction}
\algtext*{EndLoop}        % only matters if you use \Loop ... \EndLoop

\RequirePackage{comment}

\RequirePackage{cleveref}

\startlocaldefs
\usepackage{mathtools}

\usepackage[color,final]{showkeys}
\newcommand{\cut}[1]{{}}

\makeatletter
\newcommand*{\rom}[1]{\expandafter\@slowromancap\romannumeral #1@}
\makeatother

\newcommand{\cE}{{\mathcal{E}}}

\newcommand{\cL}{{\mathcal{L}}}

\newcommand{\cN}{{\mathcal{N}}}

\newcommand{\cX}{{\mathcal{X}}}

\newcommand{\cZ}{{\mathcal{Z}}}

\newcommand{\hP}{{\mathbb{P}}}

\newcommand{\bX}{{\mathbf{X}}}
\newcommand{\bY}{{\mathbf{Y}}}

 \usepackage[bb=boondox]{mathalfa}

\usepackage{xparse}
\DeclareFontFamily{U}{ntxmia}{}
\DeclareFontShape{U}{ntxmia}{m}{it}{<-> ntxmia }{}
\DeclareFontShape{U}{ntxmia}{b}{it}{<-> ntxbmia }{}
\DeclareSymbolFont{lettersA}{U}{ntxmia}{m}{it}
\SetSymbolFont{lettersA}{bold}{U}{ntxmia}{b}{it}

\newcommand{\diag}{{\mathrm{diag}}} % matrix diagonal -> vector
\makeatletter
\DeclareFontFamily{U}{tipa}{}
\DeclareFontShape{U}{tipa}{m}{n}{<->tipa10}{}
\newcommand{\arc@char}{{\usefont{U}{tipa}{m}{n}\symbol{62}}}%

\newcommand{\arc}[1]{\mathpalette\arc@arc{#1}}

\newcommand{\arc@arc}[2]{%
  \sbox0{$\m@th#1#2$}%
  \vbox{
    \hbox{\resizebox{\wd0}{\height}{\arc@char}}
    \nointerlineskip
    \box0
  }%
}
\makeatother

\definecolor{lightgrey}{gray}{0.8}
\definecolor{medgrey}{gray}{0.6}
\definecolor{darkgrey}{gray}{0.4}
\newcommand{\opI}{{\varmathbb{I}}}

\DeclarePairedDelimiter{\norm}{\lVert}{\rVert}
\DeclarePairedDelimiter{\abs}{\lvert}{\rvert}

\makeatletter
\newlength{\doublefracgap}
\DeclareRobustCommand{\doublefrac}[2]{%
  \mathinner{\mathpalette\doublefrac@{{#1}{#2}}}%
}
\newcommand{\doublefrac@}[2]{\doublefrac@@#1#2}
\newcommand{\doublefrac@@}[3]{%
  \ooalign{%
    \raisebox{\doublefracgap}{$\m@th#1\frac{#2}{\phantom{#3}}$}\cr
    \raisebox{-\doublefracgap}{$\m@th#1\frac{\phantom{#2}}{#3}$}\cr
  }%
}

\makeatother

\newcommand{\R}{\mathbb{R}}

\newcommand{\diffang}[4]{%
  	\begingroup % keep macro definitions local
		\def\firstop{#1}%
		\def\secondop{#2}%
		\def\identity{\opI}% 
		\ifx\firstop\identity
		\def\firstop{}%
		\fi
		\ifx\secondop\identity
		\def\secondop{}%
		\fi
		\angle (\firstop #3 - \firstop #4, \secondop #3 - \secondop #4)
  	\endgroup
}
\newcommand{\diffnormfrac}[4]{%
  	\begingroup % keep macro definitions local
		\def\firstop{#1}%
		\def\secondop{#2}%
		\def\identity{\opI}%
		\ifx\firstop\identity
		\def\firstop{}%
		\fi
		\ifx\secondop\identity
		\def\secondop{}%
		\fi
		\frac{\norm{\firstop #3 - \firstop #4}}{\norm{\secondop #3 - \secondop #4}}
  	\endgroup
}

\newcommand{\red}[1]{\textcolor{red}{#1}}

\newcommand{\gammatilde}{\widetilde{\gamma}}
\newcommand{\hatgammatilde}{\widehat{\widetilde{\gamma}}}

\endlocaldefs

\begin{document}

\begin{frontmatter}

\title{Model-Agnostic Search for New Physics in the Presence of Imperfect Background by Smearing the Data}
\runtitle{Model-Agnostic Search by Smearing the Data}

\begin{aug}
\author[A]{\fnms{Soheun}~\snm{Yi}\ead[label=e1]{soheuny@andrew.cmu.edu}}
\author[B]{\fnms{John}~\snm{Alison}\ead[label=e2]{johnalison@cmu.edu}}
\author[A]{\fnms{Mikael}~\snm{Kuusela}\ead[label=e3]{mkuusela@andrew.cmu.edu}}

\address[A]{Department of Statistics and Data Science,
Carnegie Mellon University\printead[presep={,\ }]{e1,e3}}

\address[B]{Department of Physics,
Carnegie Mellon University\printead[presep={,\ }]{e2}}
\end{aug}

\begin{abstract}
  In the search for new particles in high-energy physics, it is necessary to first define the signal region (SR) enriched in signal events, and then compare events in this region with the expected background distribution.
  While most existing search methods rely on prior domain knowledge of either the signal or the background, there has been less effort to relax both assumptions simultaneously, as methods that weaken one assumption typically rely on strong knowledge of the other.
  We propose \textbf{M}odel-\textbf{A}gnostic \textbf{S}earch with \textbf{S}mearing and \textbf{I}mperfect \textbf{B}ackground (MASSIB), a method that addresses this challenge by learning the density ratio of events relative to a misspecified background distribution using modern tools from machine learning, and defining the SR as the region with high density ratio.
  Recognizing that a na\"ive application of this idea introduces significant bias in background estimation when interpolating from the control region to the SR, MASSIB mitigates this issue by introducing a ``smeared'' density ratio, which can be learned efficiently through data-level manipulation and used to define an SR that allows for accurate background estimation.
  By applying it to various types of simulated Large Hadron Collider (LHC) events with realistic background mismodeling, we demonstrate that MASSIB identifies a data-driven SR that achieves high signal concentration consistently across different types of signals while avoiding bias in background estimation. The method does this without making strong assumptions about either the signal or background distribution.
  We then show that this robustness extends to the validity and strong statistical power of the resulting hypothesis tests for signal detection, demonstrating the potential of this approach for future model-agnostic searches at the LHC.
\end{abstract}

%% Uncomment and fill in the MSC codes before submission.
%\begin{keyword}[class=MSC]
%\kwd[Primary ]{62P35}
%\kwd[; secondary ]{62G10}
%\end{keyword}


\end{frontmatter}

%%%%%%%%%%%%%%%%%%%%%%%%%%%%%%%%%%%%%%%%%%%%%%
%% Please use \tableofcontents for articles %%
%% with 50 pages and more                   %%
%%%%%%%%%%%%%%%%%%%%%%%%%%%%%%%%%%%%%%%%%%%%%%
%\tableofcontents

\section{Introduction}
\label{s:intro}

High-energy physics studies fundamental particles and their interactions. The Standard Model (SM) describes our current understanding of particle physics. The discovery of the Higgs boson by collaborations at the European Organization for Nuclear Research (CERN) \citep{ATLAS2012_observation,CMS2012_observation} and its inclusion in the SM lead to a natural question about the possibility of new particles and interactions beyond the SM.

At the Large Hadron Collider (LHC), beams of protons or heavy ions are accelerated to relativistic energies and brought into collisions, which are captured by one of the detectors.
In each collision, referred to as an \emph{event}, part of the available kinetic energy is converted into mass, producing a spray of short-lived and stable particles that are subsequently reconstructed. 
The vast majority of events originate from well-understood SM processes, collectively constituting the \emph{background}. 
Far more rarely, collisions may produce particles or interactions beyond the Standard Model, giving rise to \emph{signal} events.
The core challenge of collider-based searches for new physics is to detect the presence of these rare signal events amid an overwhelming background.

From a statistical perspective, searches for new physics are hypothesis testing problems at their core.
The observed events are modeled as being drawn from a mixture of the background distribution $p_b$ and a potential signal distribution $p_s$, yielding an overall distribution $(1 - \epsilon) p_b + \epsilon p_s$, where $\epsilon$ denotes the fraction of the signal process. 
We refer to this mixing proportion as the \emph{signal ratio}.
The null hypothesis $H_0 \colon \epsilon = 0$ corresponds to a background-only scenario, while the alternative $H_1 \colon \epsilon > 0$ indicates the presence of signal events.
In the seminal Higgs boson detection analyses \citep{ATLAS2012_observation,CMS2012_observation}, this framework was implemented using detailed Monte Carlo simulations for the signal processes and channel-specific background estimates, which were then combined to construct likelihood-ratio test statistics for comparing the two hypotheses.

This approach breaks down when samples drawn from the true signal process are unavailable, as the signal distribution cannot be reconstructed.
A common strategy in this setting is to introduce a signal-specific \emph{signal region} (SR) and a corresponding \emph{control region} (CR) to be enriched and depleted in signal events, respectively, based on prior knowledge about the kinematic properties of the signal.
The background distribution is then estimated in the CR and subsequently interpolated into the SR for comparison with the observed data. 
However, the applicability of this strategy depends heavily on the ability to accurately define a signal-rich SR, which often requires identifying a small number of variables where a resonant signal is expected to manifest itself. 
While ANODE \citep{NachmanShih2020_anomalya}, CATHODE \citep{Hallin2022_CATHODE,Hallin2023_resonant}, and related methods \citep{ManoleBryantAlisonKuuselaWasserman2023_background} construct SRs based on this approach, these methods assume a degree of prior knowledge that may not be available for a completely new type of particle or interaction.

Another line of research tackles this problem by assuming direct access to samples from the true background distribution, which allows for a direct comparison between the observed and background events via two-sample testing.
The NPLM method of \citet{DAngnolo2019_LNPM} and its variants \citep{DAgnoloGrossoPieriniWulzerZanetti2021_learning,DAgnoloGrossoPieriniWulzerZanetti2022_learning} estimate the likelihood ratio between the observed data and background samples, and then apply a statistical test to assess the significance of deviations from the reference distribution.
Similarly, \citet{ChakravartiKuuselaLeiWasserman2023_modelindependent} construct test statistics using a classifier trained to distinguish between background and observed events.
However, these approaches face limitations in realistic settings, where the background is often only partially known or misspecified, rendering direct comparisons unreliable.
The search for double Higgs boson production in the four-$b$-jet final state \citep{ATLAS_HH4b,CMS_HH4b} exemplifies this challenge: the background cannot be simulated accurately and must instead be estimated from data before being compared with the observations to test for a potential signal.

\begin{figure}[t]
  \centering
  \hspace{5em}
  \includegraphics[width=0.7\textwidth]{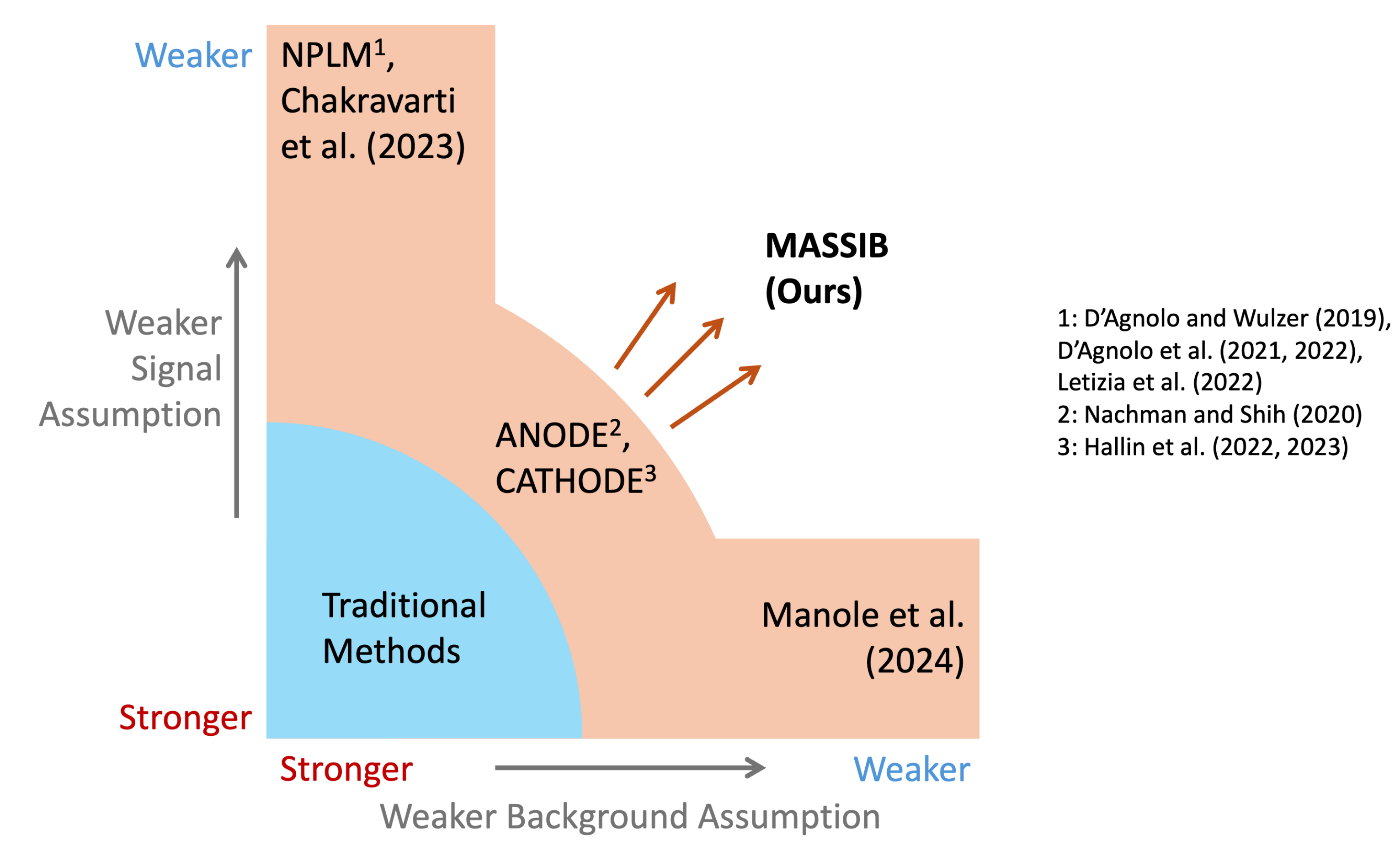}
  \caption{Classification of methods by how strong their assumptions on the signal and background are. 
  Traditional methods, including Higgs boson detection \citep{ATLAS2012_observation,CMS2012_observation}, rely on strong assumptions about both the signal and background. Some recent methods relax the assumptions on either the signal \citep{ChakravartiKuuselaLeiWasserman2023_modelindependent} or the background \citep{ManoleBryantAlisonKuuselaWasserman2023_background}, but still impose a strong assumption on the other component.
  In contrast, this work introduces a new approach built upon weaker assumptions on both the signal and background.
  }  
  \label{fig:contribution}
\end{figure}

In this work, we take a step toward simultaneously relaxing both of the assumptions by proposing \textbf{MASSIB}: \textbf{M}odel-\textbf{A}gnostic \textbf{S}earch with \textbf{S}mearing and \textbf{I}mperfect \textbf{B}ackground.
As illustrated in \Cref{fig:contribution}, our method reduces the dependence on specific prior knowledge about both the signal and background events. 
It constructs an SR that is model-agnostic to the signal distribution and estimates the background distribution in the SR by leveraging auxiliary events from an imperfect background, which share similar kinematics with the true background but cannot substitute for it directly.

The conceptual foundation of our method is the ABCD method \citep{PhysRevD.44.29,10.5555/2564838}, which is analogous to mark-recapture estimation in ecology \citep{Cren1965ANO}.
This approach organizes events in a contingency table (\Cref{fig:abcd}) comprising four regions $A$, $B$, $C$, and $D$ based on two criteria: whether an event belongs to the SR or CR, and whether it originates from the imperfect background or target samples.
Since no signal events arise from the imperfect background, the expected number of background events in region $D$ can be estimated as $N_D \approx N_B N_C / N_A$.
In the mark-recapture context, the first and second ``captures'' correspond to the horizontal and vertical axes, respectively. 
The validity of this estimate depends on the fundamental assumption that the two axes are uncorrelated.
Consequently, the key to obtaining an accurate background estimate lies in constructing the SR and CR in a way that they remain uncorrelated with the event source.

\begin{figure}
\begin{tikzpicture}[scale=1.0]
\def\width{4.5}   % total width
\def\height{2}   % total height

% Draw outer rectangle
\draw (0,0) rectangle (\width,\height);

% Vertical and horizontal lines
\draw (\width / 2,0) -- (\width / 2,\height);
\draw (0,\height/2) -- (\width,\height/2);

% Column headers
\node at (\width * 0.25, \height + 0.5) {\textbf{CR}};
\node at (\width * 0.75, \height + 0.5) {\textbf{SR}};

% Row labels
\node[anchor=east] at (-0.2,\height * 0.75) {Imperfect Background};
\node[anchor=east] at (-0.2,\height * 0.25) {Target Events};

% Cell labels
\node at (\width *0.25, \height * 0.75) {$N_A$};
\node at (\width * 0.75, \height * 0.75) {$N_B$};
\node at (\width * 0.25, \height * 0.25) {$N_C$};
\node at (\width * 0.75, \height * 0.25) {$N_D$};

\end{tikzpicture}
\hspace{2.9cm}
\caption{A contingency table explaining the ABCD method.
The four regions $A$, $B$, $C$, and $D$ are defined based on whether an event belongs to the imperfect background or the target events and whether it is in the SR or CR.
Given that the imperfect background does not include any signal events and that the two axes are uncorrelated, we can estimate the number of background events in region $D$ by $N_D \approx N_B N_C / N_A$.}
\label{fig:abcd}
\end{figure}

\begin{figure}
  \centering
  \includegraphics[width=0.95\textwidth]{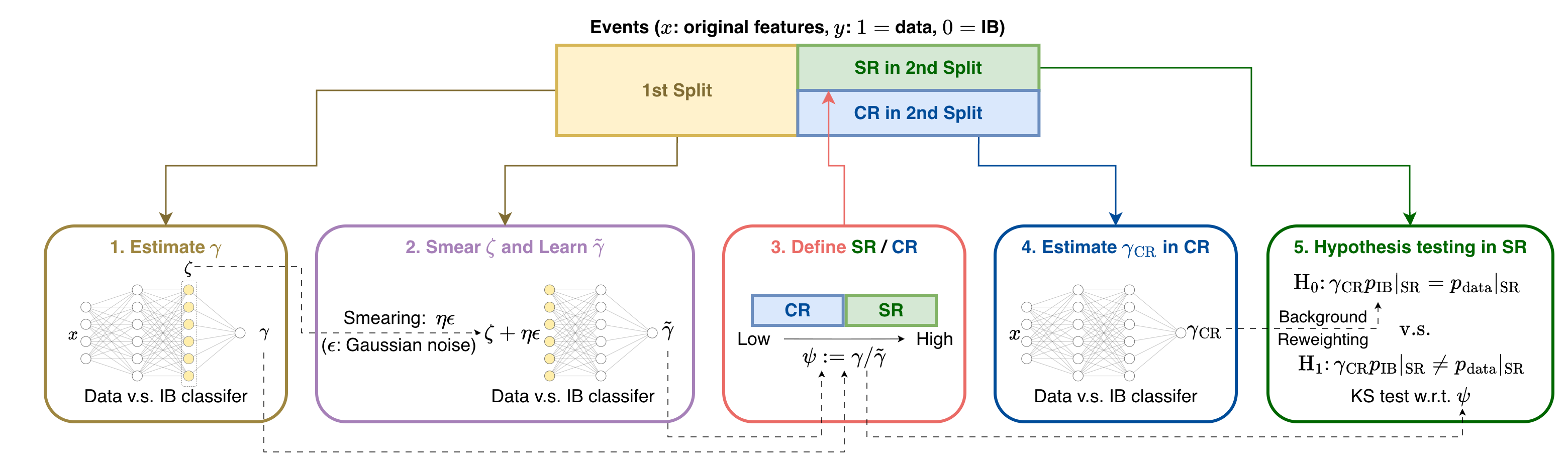}
  % \hspace{5em}
  \caption{An overview of MASSIB. The procedure begins by estimating the density ratio $\gamma$ between the data and imperfect background (IB) in the first split of the dataset. Then, the representation $\zeta$ is smeared and used to learn the density ratio $\gammatilde$ between the smeared data and smeared IB in the same split. 
  The SR and CR are then defined by thresholding on $\psi = \gamma / \gammatilde$.
  Next, the density ratio of the data and IB in the CR, denoted as $\gamma_{\text{CR}}$, is estimated.
  Finally, the estimated background $\gamma_{\text{CR}} p_{\text{IB}}$ is compared with the data distribution $p_{\text{data}}$ within the SR via hypothesis tests.
  }
  \label{fig:workflow}
\end{figure}

To achieve this, we first demonstrate that selecting the SR as a region with a high density ratio of observed events to the misspecified background leads to strong signal concentration in the SR.
However, we find that this definition of SR and CR inherently introduces the correlation problem described above, leading to an unreliable interpolation of the density ratio from the CR to the SR, resulting in significant bias in background estimation in the SR.
To overcome this, we introduce a \emph{smeared} density ratio, learned by adding noise to event representations, which decorrelates the two axes of the contingency table (\Cref{fig:abcd}).
This approach enables high signal concentration in the SR while also mitigating bias by improving the accuracy of background estimation.
An overview of our approach is provided in \Cref{fig:workflow}.

The rest of the paper is organized as follows. In \Cref{s:related_work}, we review related work in the search for new physics and its relevance to our work.
In \Cref{s:problem_formulation}, we introduce the datasets and the problem of new physics detection, and define various important concepts.
In \Cref{s:method}, we systematically explain our approach to defining the SR and CR and develop a hypothesis test, together with motivations. In \Cref{s:results}, we demonstrate the performance of our method through experiments.
\Cref{ss:res_concentration_of_signal_events_in_the_SR} demonstrates concentration of signal events in the SR, \Cref{ss:res_hypothesis_testing} displays the validity and power of hypothesis tests based on our SR selection, and \Cref{ss:res_generalization} demonstrates generalization to different signal types. In \Cref{s:discussion}, we discuss the implications of our method and potential future work.

\section{Related Work}
\label{s:related_work}

The idea of defining a CR or sideband to estimate the background and interpolate it into an SR underlies many methods for new physics discovery. 
These approaches generally assume that the signal events are localized in an appropriate feature space, often manifesting as a high-frequency bump, which naturally defines an SR and CR either deterministically or through a scanning procedure.
The traditional ABCD method 
\citep{PhysRevD.44.29,10.5555/2564838} is among the earliest implementations of this concept in high-energy physics.

The advent of modern machine learning (ML) has significantly advanced this framework through powerful data-driven techniques; see \citet{AmramLetiziaKuusela2026modelagnostic} for a recent review. 
These developments span weakly supervised, unsupervised, and hybrid approaches, reflecting different degrees of event-level supervision. 
Weakly supervised methods include the CWoLa framework \citep{MetodievNachmanThaler2017_classification,CollinsHoweNachman2018_anomaly,CollinsHoweNachman2019_extending} and the Tag N' Train method \citep{AmramSuarez2021_tag}. 
CWoLa trains a classifier on two mixed samples with different signal fractions, which yields the optimal signal-to-background classifier that can be applied for signal detection.
The Tag N' Train method iteratively tags events as signal-like and background-like by a weak classifier, then refines the labeling with a stronger classifier.
CATHODE \citep{Hallin2022_CATHODE,Hallin2023_resonant} uses normalizing flows to learn the background distribution in the CR, samples synthetic background events in the SR, and trains a classifier to distinguish the observed data from those samples.
The CURTAINs framework \citep{RaineKleinSenguptaGolling2023_curtains,Sengupta2024_curtains} follows the same pattern, ``transporting'' CR events into the SR with invertible neural networks before training the classifier.
Unsupervised methods include ANODE \citep{NachmanShih2020_anomalya}, which employs a neural density estimator to learn the background distribution in the CR, interpolates it into the SR, and uses the resulting density ratio directly as an anomaly score.
Several autoencoder-based methods \citep{FarinaNakaiShih2020_autoencoder,HeimelKasieczkaPlehnThompson2019_qcd} have also been developed for anomaly detection in an unsupervised manner.
Hybrid approaches leverage simulations to augment data-driven searches for new physics.
For instance, the FETA approach \citep{Golling2023_feta} extends the CURTAINs framework by involving simulated data, and the SALAD method \citep{AndreassenNachmanShih2020_simulation,BenkendorferLePottierNachman2021_simulationassisted} learns a density ratio of simulation to data in sidebands, then interpolates this into the SR for background estimation.

The methods most closely related to MASSIB can be compared along two dimensions: how accurately the background distribution must be specified and how much signal-specific information is needed to define the SR. 
Several methods reviewed above reduce dependence on an exact background model by using information from signal-poor sidebands to estimate the background in the SR. 
These approaches generally require a prespecified resonant variable to define the SR and CR and assume that the relationship learned from the sidebands remains valid in the SR; as \citet{Kasieczka2026_multijet} note, they have so far relied on a resonance decaying to a pair of jets.
The four-$b$-jet channel considered here offers neither: the four jets admit three distinct pairings into Higgs candidates, so no single bump-hunting variable is well defined, and non-resonant $\mathrm{HH} \to 4b$ production has no unknown resonance to scan over.
Within this literature, \citet{ManoleBryantAlisonKuuselaWasserman2023_background} study the related problem of constructing the target background from a structurally different auxiliary background sample using density-ratio estimation or optimal transport.
Like the preceding approaches, their methods transfer information learned in the CR into the SR, while retaining a signal-specific definition of the SR and CR.

A complementary line of work relaxes the signal-side requirement while relying on a reference distribution that closely approximates the background. 
The NPLM approach \citep{DAngnolo2019_LNPM} trains a neural network to learn the likelihood ratio between observations and a reference distribution, enabling signal-model-agnostic anomaly detection. 
Follow-up work extends this framework to related settings \citep{DAgnoloGrossoPieriniWulzerZanetti2021_learning,LetiziaLosapioRandoGrossoWulzerPieriniZanettiRosasco2022_learning}, including the treatment of model uncertainty through nuisance parameters \citep{DAgnoloGrossoPieriniWulzerZanetti2022_learning}. 
More recently, \citet{ChakravartiKuuselaLeiWasserman2023_modelindependent} proposed a model-independent procedure that tests for the presence of signal using a classifier-based two-sample test \citep{KimLeeLei2019_global,KimRamdasSinghWasserman2021_classification}, without partitioning the data into an SR and CR.

MASSIB combines these two directions by using a misspecified auxiliary background while constructing the SR and CR from data.
It therefore requires limited prior information about both the signal and background.
On the signal side, MASSIB requires only a signal-concentration condition: signal events manifest as a localized, high-frequency feature in the learned representation relative to the more gradually varying discrepancy between the auxiliary and target background distributions.
Under this condition, MASSIB does not require detailed signal-specific information, such as a prespecified resonant variable.

% For a comprehensive overview of related approaches that leverage ML for new physics discovery, we refer the reader to \citep{LHCO2021_lhc}.

% Recent searchs: \citep{CMS2025_modelagnostic,ATLAS2025_weakly,CMS:2025dbd}

% NPLM: related? 
% CWoLa: \citep{MetodievNachmanThaler2017_classification}
% CWoLa Hunting: \citep{CollinsHoweNachman2018_anomaly,CollinsHoweNachman2019_extending}

% Tag N' Train: \citep{AmramSuarez2021_tag}
% CATHODE: \citep{Hallin2022_CATHODE,Hallin2023_resonant}
% QUAK: \citep{ParkRankinUdrescuYunusHarris2021_quasi},
% SALAD: \citep{AndreassenNachmanShih2020_simulation,BenkendorferLePottierNachman2021_simulationassisted}
% CURTAINS: \citep{RaineKleinSenguptaGolling2023_curtains,Sengupta2024_curtains}
% FETA: \citep{Golling2023_feta}
% Auto Encoders: \citep{HeimelKasieczkaPlehnThompson2019_qcd,FarinaNakaiShih2020_autoencoder}

% Two sample testing as classifier is related to \citep{KimLeeLei2019_global,KimRamdasSinghWasserman2021_classification}.

\section{Problem Formulation}
\label{s:problem_formulation}

When two protons collide, they typically form unstable particles that decay within an extremely short time frame, making them impossible to detect directly. Some of these transient particles decay into quarks, which interact with the vacuum to produce collimated sprays of stable particles, referred to as ``jets''. Following the setting of \citet{ManoleBryantAlisonKuuselaWasserman2023_background}, we focus on events resulting in four jets.

Being able to detect jets efficiently, the Compact Muon Solenoid (CMS) detector at CERN records the momentum and invariant mass (or energy) of the particles produced in the proton-proton collisions.
The momentum vector is represented by a three-dimensional vector $(p_T, \eta, \phi)$, where $p_T$ is the \emph{transverse momentum}, or the magnitude of the momentum projected onto the $xy$-plane, $\phi$ is the azimuthal angle, and $\eta = -\log \tan(\theta / 2)$ is the \emph{pseudorapidity} characterized by the polar angle $\theta$.
Writing the rest mass of a particle as $m$, the measured features of an observed object are represented as a four-vector $(p_T, \eta, \phi, m) \in \R^4$. As we are concerned with collisions resulting in $4$ jets, the corresponding events can be represented with $x = (p_T^i, \eta^i, \phi^i, m^i)_{i=1}^4 \in \R^{16}$, forming a $16$-dimensional feature space $\cX \subset \R^{16}$.

Events in which four jets are \emph{tagged} as originating from $b$-quarks are referred to as $4b$ events. 
These events are predominantly \textbf{background events} predicted by the SM and arise mainly from quantum chromodynamics (QCD) processes, which makes it hard to simulate them accurately using theoretical models.
In addition to the $4b$ events, we also consider events with four jets where only three of them are $b$-tagged: we refer to these events as $3b$ events.
The $3b$ events serve as an imperfect version of the background $4b$ events: although they share similar kinematic properties, they cannot fully replace the $4b$ background.
We leverage the density ratio of $4b$ events to $3b$ events to define the SR and CR, which we further elaborate on in \Cref{s:method}.

Our objective is to detect potential \textbf{signal events} among $4b$ events, which constitute a very small fraction and are not part of the SM background.
A prominent example of signal events in this channel is the process $\mathrm{HH} \to 4b$, in which two Higgs bosons are produced and each decays into a pair of $b$-quarks, yielding four $b$-tagged jets.
When two Higgs bosons originate from the decay of a single heavy particle, we refer to this as a \emph{resonant} signal and denote the process by $X \to \mathrm{HH} \to 4b$.
In contrast, \emph{non-resonant} $\mathrm{HH} \to 4b$ events involve direct Higgs-pair production without an intermediate resonance.
Another example of a potential signal is non-resonant $\mathrm{ZH} \to 4b$, where a $\mathrm{Z}$-boson and a Higgs boson are produced together and both decay into two $b$-tagged jets, resulting again in a $4b$ final state.

Because proton-proton collisions occur independently, the resulting event samples are naturally modeled as Poisson point processes \citep{Reiss1993_course,ManoleBryantAlisonKuuselaWasserman2023_background}. Specifically, we model the $3b$ background, $4b$ background, and signal event processes as mutually independent Poisson point processes on $\cX$ with intensity measures $\beta_{3b}$, $\beta_{4b}$, and $\mu\sigma$, respectively, where $\mu \geq 0$ represents the strength of the signal process. Thus, for each process, the number of events in any measurable region $A \subseteq \cX$ is Poisson with mean equal to its intensity measure on $A$, and counts in disjoint regions are mutually independent. We assume that $\beta_{3b}$, $\beta_{4b}$, and $\sigma$ admit intensity functions $b_{3b}$, $b_{4b}$, and $s$, respectively.

Since the $4b$ sample is a superposition of background $4b$ and signal events, the observed $4b$ event process is a Poisson point process with intensity measure $\lambda_{4b} \coloneqq \beta_{4b} + \mu \sigma$ and intensity function $\ell_{4b} \coloneqq b_{4b} + \mu s$. We have access to ${n_{3b}}$ samples of $3b$ and ${n_{4b}}$ samples of $4b$ events, denoted by $\{X_{3b}^i\}_{i=1}^{n_{3b}}$ and $\{X_{4b}^j\}_{j=1}^{n_{4b}}$, respectively. The Poisson point-process assumptions imply that, conditional on ${n_{3b}}$ and ${n_{4b}}$,
\begin{equation}
    \begin{aligned}
        X_{3b}^1, \dots, X_{3b}^{n_{3b}} \stackrel{\text{i.i.d.}}{\sim} P_{3b} 
        &\coloneqq \frac{\beta_{3b}}{\beta_{3b}(\cX)}, \\
        X_{4b}^1, \dots, X_{4b}^{n_{4b}} \stackrel{\text{i.i.d.}}{\sim} P_{4b} 
        &\coloneqq \frac{\lambda_{4b}}{\lambda_{4b}(\cX)} 
        = (1 - \epsilon) B_{4b} + \epsilon S,
    \end{aligned}
\end{equation}
where $B_{4b} \coloneqq \beta_{4b} / \beta_{4b}(\cX)$, $S \coloneqq \sigma / \sigma(\cX)$, and $\epsilon \coloneqq \mu \sigma(\cX) / \lambda_{4b}(\cX)$. We call $\epsilon$ the \emph{signal ratio}, which is the expected fraction of signal events among the $4b$ events.

Since a nonzero $\epsilon$ indicates the existence of a signal in the data, the problem of discovering new physics boils down to testing the following hypotheses:
\begin{equation}\label{eq:sr_ratio_test}
    H_0 \colon \epsilon = 0 \quad \text{vs.} \quad H_1 \colon \epsilon > 0.
\end{equation}
Each simulated event carries a weight reflecting the combinatorial and kinematic reweighting applied to the samples.
Although $3b$ events are more abundant in the underlying data, we balance the $3b$ and $4b$ samples by total event weight in our experiments. Throughout this paper, we therefore work under the convention $\beta_{3b}(\cX)=\lambda_{4b}(\cX)$. Under this convention, the mixture induced by the $3b$ and $4b$ event processes has class probabilities
$
\hP(Y=0)=\hP(Y=1)=\frac{1}{2},
$
where $Y=0$ denotes a $3b$ event and $Y=1$ denotes a $4b$ event. We define the \emph{density ratio} of $4b$ events to $3b$ events as
\begin{equation}\label{eq:density_ratio}
    \gamma(x)
    \coloneqq
    \frac{\hP(Y=1\mid X=x)}{\hP(Y=0\mid X=x)}
    =
    \frac{P_{4b}(x)}{P_{3b}(x)}
    =
    \frac{b_{4b}(x)+\mu s(x)}{b_{3b}(x)}
    =
    \frac{(1-\epsilon)B_{4b}(x)+\epsilon S(x)}
         {P_{3b}(x)},
\end{equation}
where we overload $P_{3b}$, $P_{4b}$, $B_{4b}$, and $S$ to denote their corresponding probability density functions.
Under this balanced construction, \eqref{eq:density_ratio} shows that learning the class-probability odds is equivalent to learning the density ratio $P_{4b}/P_{3b}$, following the classifier-based likelihood-ratio estimation approach of \citet{CranmerPavezLouppe2015_likelihood}.
For clarity of exposition, the main text presents all definitions and algorithms as if every event had unit weight; \Cref{s:weights} describes these weights and states the weight-aware versions used in our implementation.

\section{Methods}
\label{s:method}

In the following sections, we will use the symbols $x$, $X$, $\bX$, and $\cX$, potentially with subscripts, to represent a specific value, a random variable, a set of random variables (or samples), and the space of observed features of the events, respectively. We also refer to $\cX$ as the \emph{original space}. Similarly, we use $y$, $Y$, and $\bY$ to represent the same concepts for the labels of the events, i.e., whether an event is $3b$ or $4b$.

\subsection{Learning Density Ratios and Representations}
\label{ss:learning_density_ratio}
We adopt the neural network architecture of \citet[Appendix B]{ManoleBryantAlisonKuuselaWasserman2023_background} to estimate the density ratio. The architecture is designed primarily around the pair structure of the relevant four-jet events: it considers all possible jet pairings and encodes the corresponding pairing-specific features in intermediate network nodes, while respecting the reflection symmetries in $\phi$ and $\eta$ and the rotational symmetry in $\phi$.
\begin{figure}
  \centering
  \begin{tikzpicture}
      % encoder - decoder tikz
      % rectangle to represent input
      \def\inputwidth{0.5}
      \def\inputheight{1}
      \def\encwidth{4}
      \def\encheight{0.5}
      \def\reprwidth{1}
      \def\decwidth{3}
      \def\decheight{0.6}
      \def\outputwidth{2}

      \draw (-\inputwidth/2, -\inputheight/2) rectangle (\inputwidth/2, \inputheight/2);
      \coordinate (inputLT) at (-\inputwidth/2, \inputheight/2);
      \coordinate (inputRT) at (\inputwidth/2, \inputheight/2);
      \coordinate (inputRB) at (\inputwidth/2, -\inputheight/2);
      \coordinate (reprLB) at ($(inputRB) + (\encwidth, -\encheight)$);
      \draw[dashed] (inputRB) -- (reprLB);
      \coordinate (reprLT) at ($(inputRT) + (\encwidth, \encheight)$);
      \coordinate (reprRT) at ($(reprLT) + (\reprwidth, 0)$);
      \draw[dashed] (inputRT) -- (reprLT);
      \coordinate (reprRT) at ($(reprLT) + (\reprwidth, 0)$);
      \draw (reprLB) rectangle (reprRT);
      \coordinate (reprRB) at ($(reprLB) + (\reprwidth, 0)$);
      \coordinate (outLB) at ($(reprRB) + (\decwidth, \decheight)$);
      \draw[dashed] (reprRB) -- (outLB);
      \coordinate (outLT) at ($(reprRT) + (\decwidth, -\decheight)$);
      \draw[dashed] (reprRT) -- (outLT);
      \coordinate (outRT) at ($(outLT) + (\outputwidth, 0)$);
      \draw (outLB) rectangle (outRT);
      \coordinate (A) at ($(inputLT)!0.5!(inputRB)$);
      \node at (A) {$x$};
      \coordinate (B) at ($(reprLB)!0.5!(reprRT)$);
      \node at (B) {$\zeta(x)$};
      \coordinate (C) at ($(outRT)!0.5!(outLB)$);
      \node at (C) {$(\widehat{\gamma} \circ \zeta)(x)$};
      \node at ($(A)!0.5!(B)$) {Encoder (\(\zeta\))};
      \node at ($(B)!0.45!(C)$) {Classifier \((\widehat{\gamma})\)};
  \end{tikzpicture}
  \caption{Abstraction of the Four-vs-Three (FvT) classifier architecture \citep{ManoleBryantAlisonKuuselaWasserman2023_background}. $x$ denotes the original input features, $\zeta = \zeta(x)$ is the representation, and $\widehat{\gamma}$ is the output of the classifier predicting the density ratio of $4b$ events to $3b$ events.
  }
  \label{fig:classifier_architecture}
\end{figure}
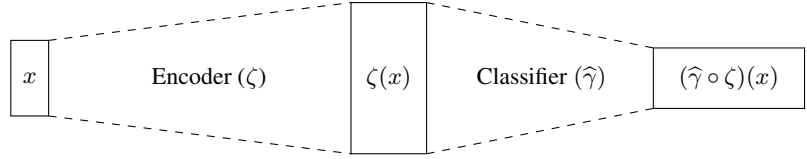

The classifier is trained to distinguish $4b$ events from imperfect background events, chosen to be \(3b\) events, as previously noted.
In particular, the classifiers are trained with a weighted loss that accounts for the event weights; see \Cref{s:weights}.
The last layer of the classifier calculates the probability of an event being a $4b$ event from inputs generated by the preceding layers. We call the inputs to the last layer the \emph{representation} of an event.
Given that the final layer is linear, the final discrimination is represented linearly in the learned representation.
We expect signal events to form a localized, high-frequency feature in this space.
The representation is trained on the contrast that carries the signal: the classifier separates $3b$ from $4b$ events and the signal lies within the $4b$ sample, so training keeps the directions along which signal events are most distinctively $4b$ and maps those events close to one another.
Such a concentration varies far more rapidly across the representation space than the diffuse discrepancy between $3b$ and background $4b$ events, which is what we mean by a high-frequency feature; \Cref{fig:tsne_original_repr} compares the two spaces and is consistent with this picture. 
We denote the representation of an event $x$ as $\zeta(x)$, or simply $\zeta$ when the context is clear. The space consisting of all possible representations is called the \emph{representation space}, defined as $\cZ \coloneqq \{\zeta(x)\,|\,x \in \cX\}$. For future reference, we formulate this process of learning the density ratio and representations as \Cref{alg:learn_density_ratio_and_representations}.
In its second procedure, $\mathbf{U}$ denotes the feature sample supplied to the classifier: $\bX_{\mathrm{CR}}$ for $\widehat{\gamma}_{\mathrm{CR}}$ and the smeared representations for $\widehat{\widetilde{\gamma}}$.

\begin{algorithm}
    \caption{Learning the density ratio and representations}
    \begin{algorithmic}[1]
        \Procedure{LearnDensityRatioAndRepresentation}{$\bX$, $\bY$}
        \State Construct a classifier $\widehat{\gamma} \circ \zeta$ as in \Cref{fig:classifier_architecture}
        \State Train the classifier to learn $\frac{\hP(Y = 1 \mid X = x)}{\hP(Y = 0 \mid X = x)}$ using $\bX$ and $\bY$
        \State \Return $\widehat{\gamma}$: estimated density ratio, $\zeta$: intermediate representation
        \EndProcedure
        \Procedure{LearnDensityRatio}{$\mathbf{U}$, $\bY$}
        \State Train a classifier $\widehat{\gamma}$ to learn $\frac{\hP(Y = 1 \mid U = u)}{\hP(Y = 0 \mid U = u)}$ using $\mathbf{U}$ and $\bY$
        \State \Return $\widehat{\gamma}$: estimated density ratio
        \EndProcedure
    \end{algorithmic}
    \label{alg:learn_density_ratio_and_representations}
\end{algorithm}

\subsection{Ensembles of Density Ratios for Signal Concentration}
\label{ss:ensemble}

Although the misspecified and true $4b$ background distributions are similar because of their shared kinematic properties, the signal distribution is expected to differ more markedly from the background distribution. Consequently, when a non-negligible fraction of signal events is present among the $4b$ events, a classifier trained on imperfect-background and $4b$ samples can assign larger density ratios to regions in which signal is concentrated. 
This motivates selecting the region with the largest density ratios as the SR.

\begin{figure}
  \centering
  \includegraphics[width=0.95\textwidth]{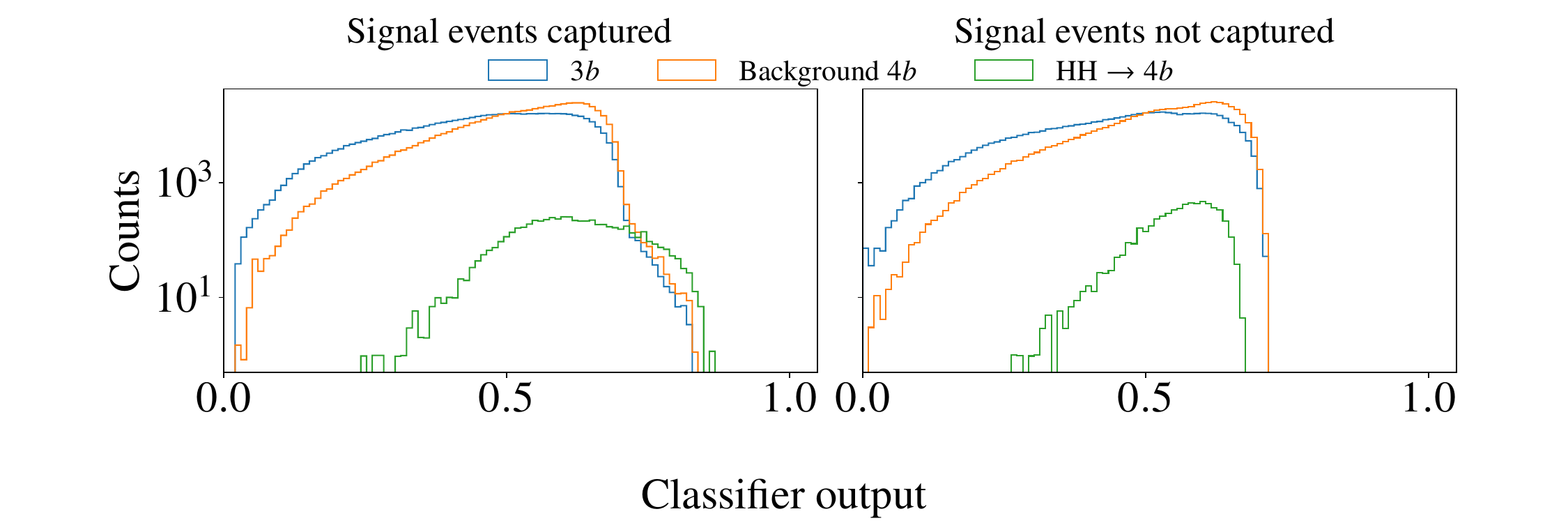}
  \caption{Histograms of classifier outputs trained on a dataset with signal ratio $\epsilon=0.01$.
  Even when trained on the same training dataset, due to the stochasticity of the training, a classifier may or may not capture signal events by identifying a region where the $\mathrm{HH} \to 4b$ signal events dominate over the background $4b$ events, as illustrated in the left and right panels.}
  \label{fig:classifier_example_hist}
\end{figure}

However, this approach faces a fundamental challenge: the dataset's signal fraction may be close to the classifier's detectable threshold. In that case, defining the SR using a single classifier may fail to concentrate enough signal events. For example, the right panel of \Cref{fig:classifier_example_hist} shows a dataset with signal ratio $\epsilon=0.01$ where selecting the SR as the classifier's high-output region does not produce a high signal fraction.

\begin{figure}[t]
  \centering
  \includegraphics[width=0.45\textwidth]{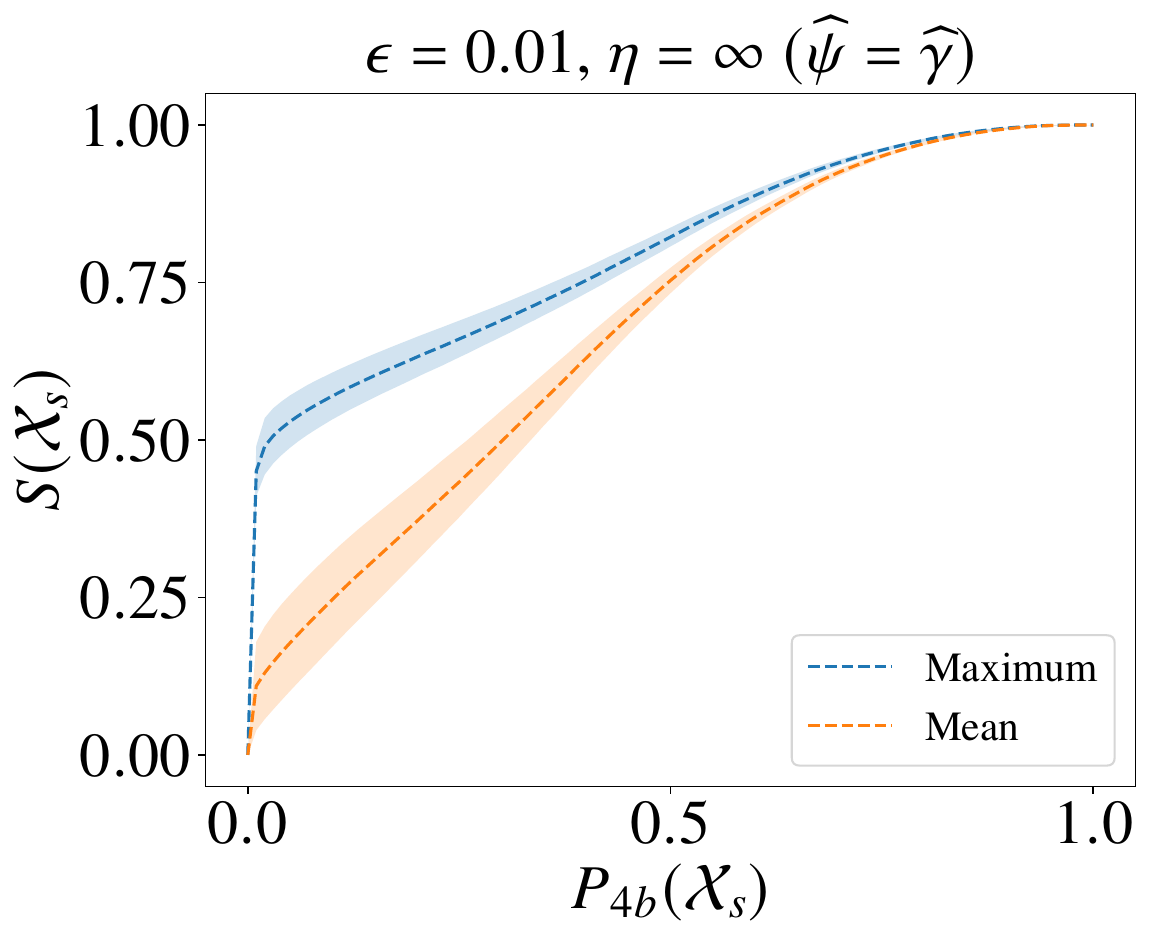}
  \includegraphics[width=0.45\textwidth]{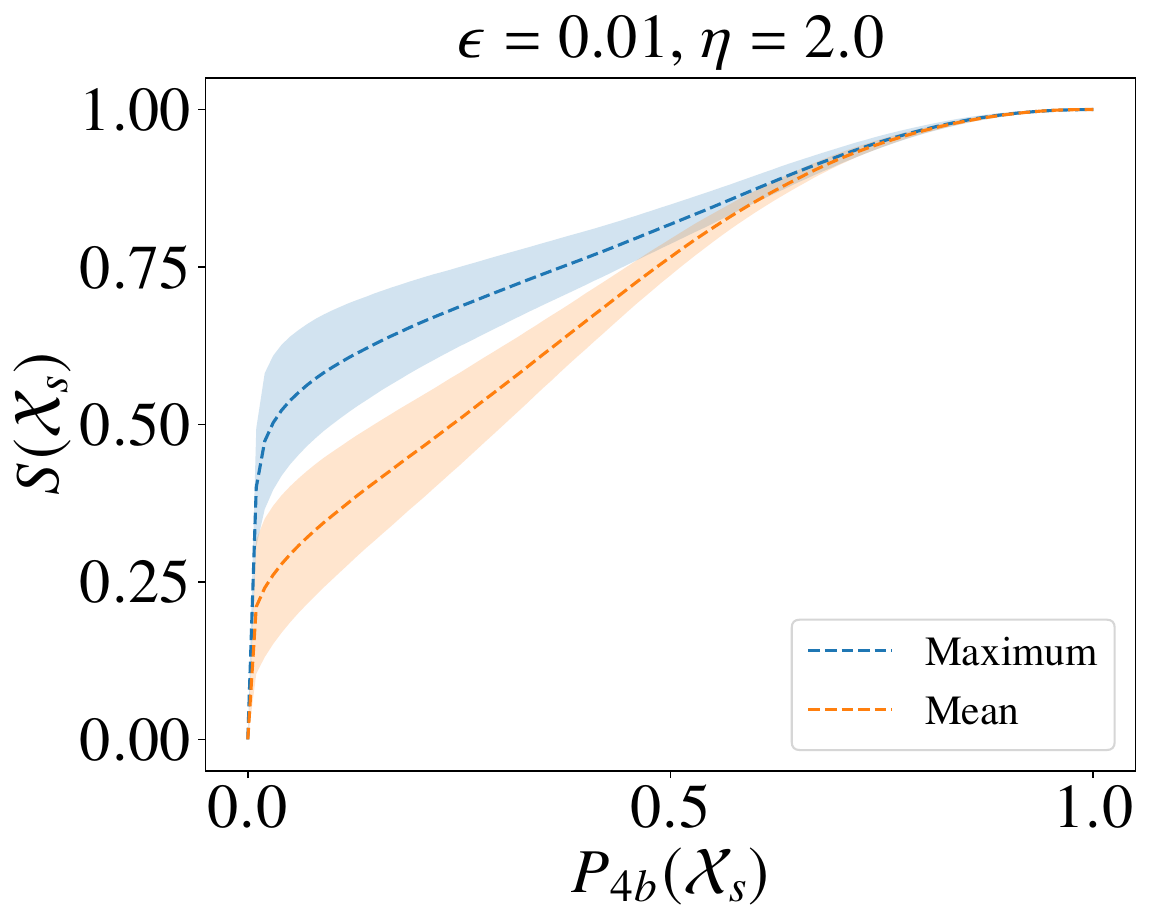}
  \caption{Proportion of signal events ($S(\cX_s)$) versus that of $4b$ events ($P_{4b}(\cX_s)$) in the SR (\(\cX_s\)) with varying thresholds for ensembles of $\widehat\psi$.
  The signal ratio is set to $\epsilon = 0.01$ and the noise scales are set to $\eta = \infty$ (i.e., \(\widehat\psi = \widehat\gamma\) is an ensemble maximum of density ratios) and \(\eta = 2.0\).
  The maximum and mean of $\widehat\psi$ for $15$ different training instances are used to define the SR.
  The plot illustrates that the ensemble maximum is more effective than the ensemble mean in enriching the signal events in the SR for both smearing and non-smearing cases.
  }
  \label{fig:ensemble_of_density_ratio}
\end{figure}

To mitigate this, we use an ensemble of classifiers to define the SR and for each event take the maximum score across the ensemble. This choice is motivated by our goal of including as many signal events as possible: if one ensemble member assigns a high density ratio to a signal event while others do not (as illustrated by the left and right panels of \Cref{fig:classifier_example_hist}), the ensemble mean would dilute that signal and exclude the event, whereas the ensemble maximum preserves the high score and gives the event a chance to enter the SR. Using the maximum therefore increases the probability that signal events are captured in the SR, yielding stronger signal enrichment in the SR. The left panel in \Cref{fig:ensemble_of_density_ratio} provides evidence that the ensemble maximum is more effective.

\subsection{Estimating Background and Using Smearing to Define SR}
\label{ss:background_estimation_strategy_and_smearing}

To obtain a data-driven background estimate in the SR, we begin by reviewing the background estimation method proposed by \citet{ManoleBryantAlisonKuuselaWasserman2023_background}, which employs reweighting of $3b$ events.
Recall that the definition of the density ratio is given by
\[
    \gamma(x) = \frac{b_{4b}(x) + \mu s(x)}{b_{3b}(x)}
\]
as in \eqref{eq:density_ratio}. 
A successful configuration of the CR ensures that it is free of signals, which can be interpreted as $s(x) \approx 0$ for $x \in \cX_c$, where $\cX_c$ denotes the CR.
Given a classifier $\widehat{\gamma}_{\text{CR}}$ trained using CR data, we can approximate $ b_{4b}(x) / b_{3b}(x) \approx \widehat{\gamma}_{\text{CR}}(x)$ in the CR.
Interpolating this classifier into the SR yields the following estimator of $b_{4b}$:
\begin{equation*}
    \widehat{\gamma}_{\text{CR}}(x) b_{3b}(x) \approx b_{4b}(x),
\end{equation*}
which effectively ``reweights'' the density of $3b$ events in the SR by using the estimated density ratio in the CR. 
We emphasize that it is crucial for $\widehat{\gamma}_{\text{CR}}$ to remain unbiased in the SR in order to reliably estimate the background distribution in the SR.

\begin{figure}
  \centering
  \includegraphics[width=0.7\textwidth]{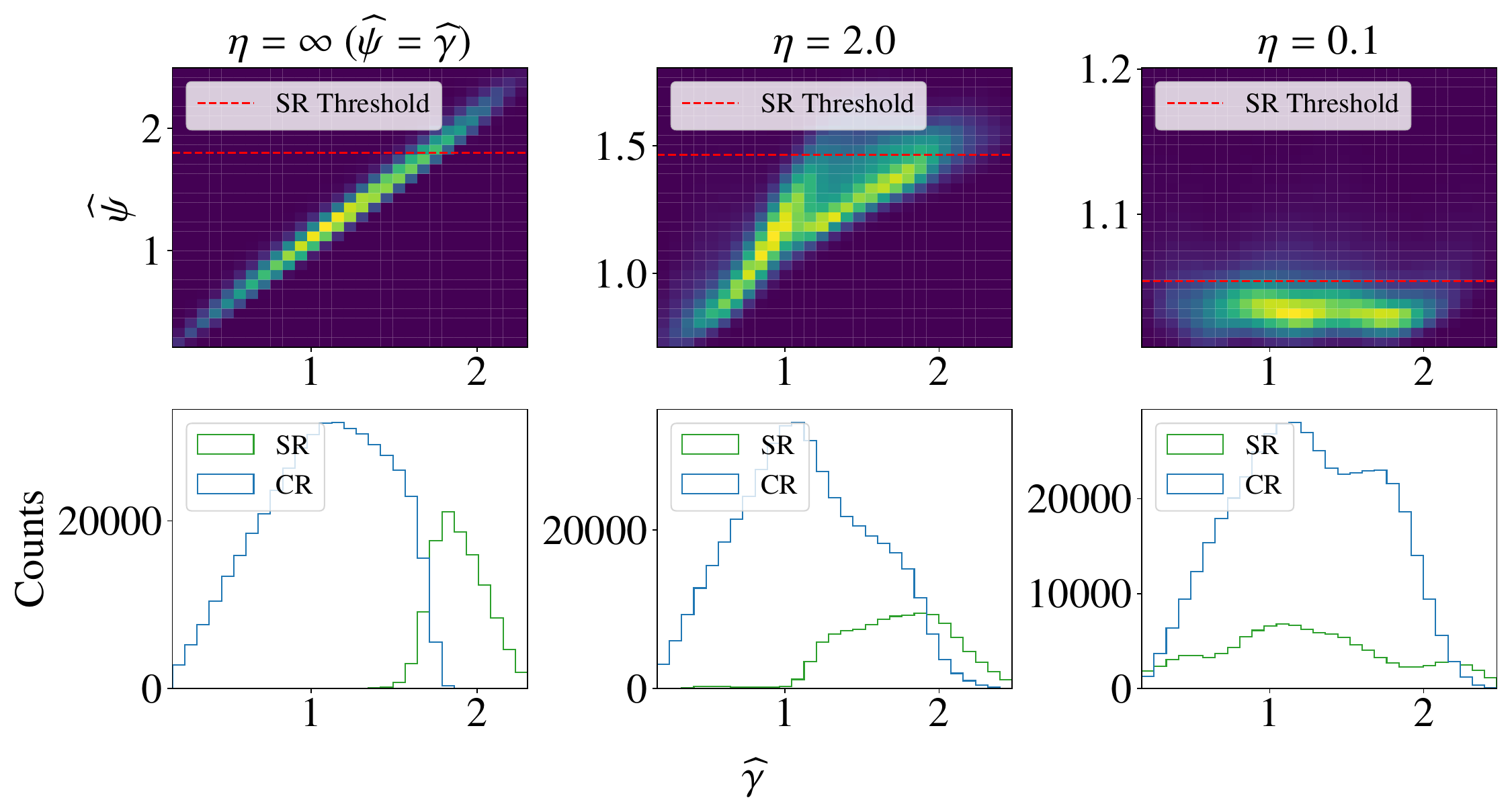}
  % \hfill
  \hspace*{0.6cm}\includegraphics[width=0.72\textwidth]{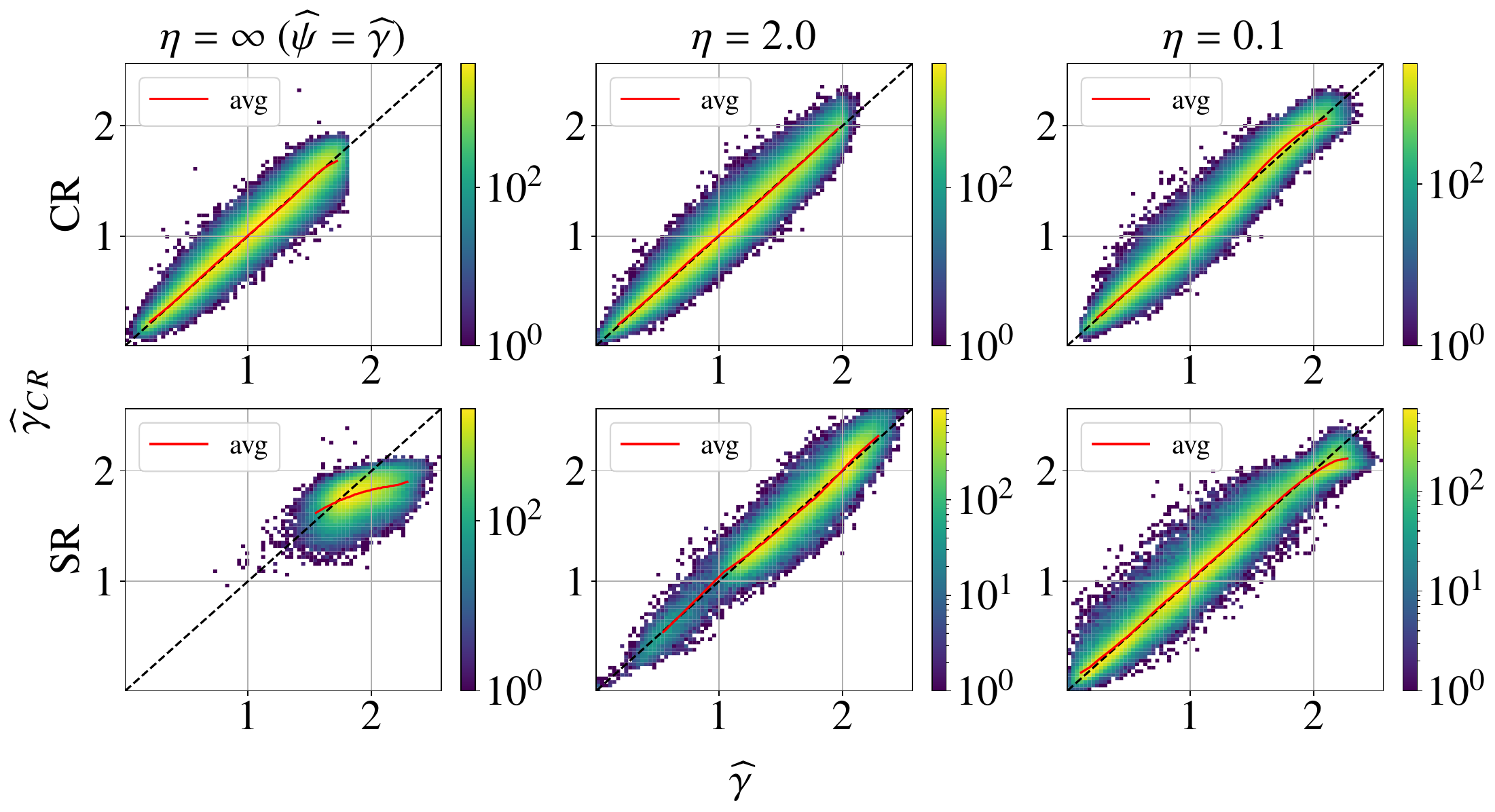}
  \caption{
    In the signal-free case (\(\epsilon=0\)), defining the SR and CR by thresholding \(\widehat{\gamma}\) reduces their overlap in terms of \(\gamma\) (first column, top two panels). 
    Consequently, a classifier trained in the CR cannot reliably learn the high values of \(\gamma\) occurring in the SR: \(\widehat{\gamma}_{\mathrm{CR}}\) is calibrated in the CR but poorly calibrated in the SR (first column, bottom panel). 
    Defining the regions by thresholding \(\widehat{\psi}\), rather than \(\widehat{\gamma}\), increases the overlap and improves calibration in the SR (second column). 
    At the smaller noise scale \(\eta=0.1\), \(\widehat{\psi}\) is close to \(1\) across most of the range of \(\widehat{\gamma}\), producing still greater overlap between the SR and CR (third column). 
    In the calibration panels, the red curve plots the event-weighted mean of \(\widehat{\gamma}_{\mathrm{CR}}\) against the event-weighted mean of \(\widehat{\gamma}\) within each of \(50\) equal-count (quantile) bins of \(\widehat{\gamma}\), and the dashed black line is the identity.
  }
  \label{fig:smearing_for_overlap_in_fvt}
\end{figure}

\begin{figure}
  \centering
  \includegraphics[width=0.9\textwidth]{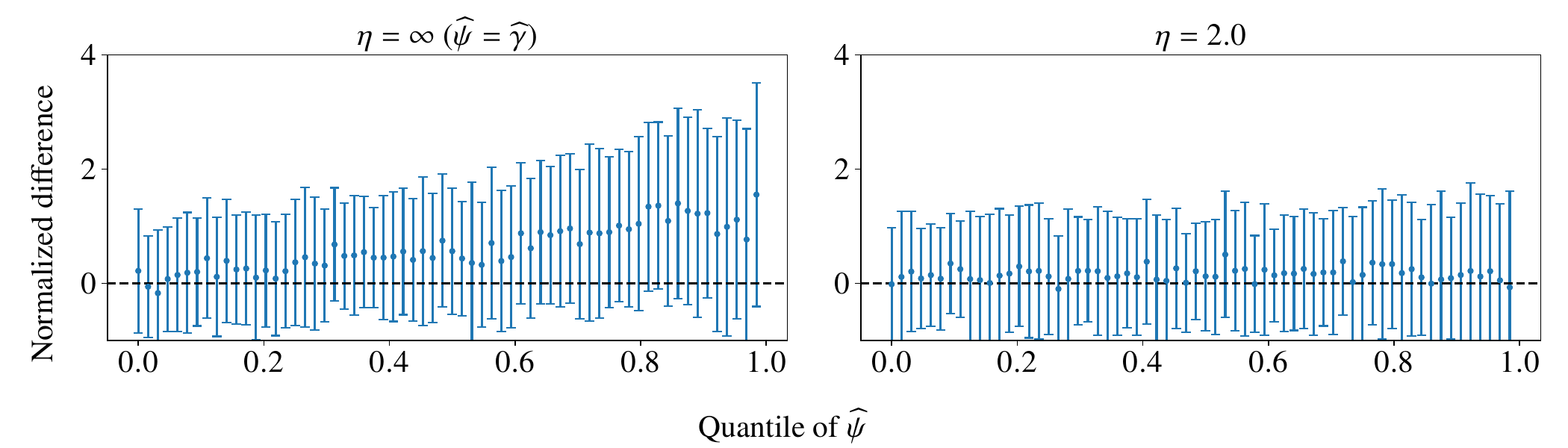}
  \caption{Quantiles of $\widehat\psi$ versus the normalized difference between the true and estimated $4b$ event counts in the SR in the signal-free case (\(\epsilon = 0\)).
  Events in the SR are binned into $64$ quantiles of $\widehat\psi$, and the mean \(\pm\) standard deviation of the normalized difference between the sum of weights of background $4b$ events and its estimate are shown across \(50\) replications.
  When \(\widehat\psi = \widehat\gamma\) (left), bias grows with $\widehat\psi$, inducing poor calibration of $\widehat\gamma_{\text{CR}}$ in the SR.
  When smearing with noise scale \(\eta = 2.0\) (right), the bias is reduced and the calibration of $\widehat\gamma_{\text{CR}}$ in the SR is improved. Here, $\eta$ denotes the noise scale used for smearing; see \Cref{alg:define_regions}.
  } 
  \label{fig:pull_vs_sr_stats}
\end{figure}

\Cref{ss:ensemble} suggests a natural approach for defining the SR: selecting regions with the highest values of the learned density ratio $\widehat{\gamma}$ derived from our ensemble of classifiers.
However, this can introduce substantial bias into the background estimation. 
An intuitive way to understand this issue is through the analogy with the ABCD method: for this approach to be valid, the variable used to define the SR and CR should be uncorrelated with \(\gamma\), the target quantity being interpolated. 
Otherwise, interpolating density ratios from the CR to the SR will lead to significant bias.

In our case, defining the SR by selecting a region with high \(\widehat{\gamma}\) and the CR as a lower-$\widehat{\gamma}$ region reduces the overlap between the SR and CR in terms of the underlying density ratio \(\gamma\), as depicted in \Cref{fig:smearing_for_overlap_in_fvt}.
With limited overlap, a classifier trained on the CR cannot capture the high density ratios of the SR.
In particular, a classifier trained only on the CR would never learn high values of \(\gamma\) and therefore would not output them in the SR. 
As a result, the density ratio $\widehat{\gamma}_{\text{CR}}$, though well calibrated on the CR, generalizes poorly to the SR and exhibits systematic miscalibration, as illustrated in \Cref{fig:pull_vs_sr_stats}.

To address this challenge, we need to define the SR and CR so that both regions have sufficient overlap in the density ratio space.
We implement this by \emph{smearing} the event representations: adding random noise with scale $\eta$ convolves the representation-space densities with the corresponding noise kernel.
We consider the signal as a relatively high-frequency feature, whereas the difference between $3b$ and background $4b$ events acts as a relatively low-frequency feature in the representation space. 
This assumption, as discussed earlier in \Cref{ss:learning_density_ratio}, is based on the localized nature of signal events in contrast to the more gradually varying background differences.
Building on the insights derived from the frequency characteristics, we are motivated by the concept of a low-pass filter.
This involves extracting low-frequency components from both the $4b$ and $3b$ densities and subsequently comparing the resulting density ratio with the original to detect high-frequency components.

\begin{figure}
  \centering
  \includegraphics[width=0.48\textwidth, trim={0 -0.3cm 0 0}, clip]{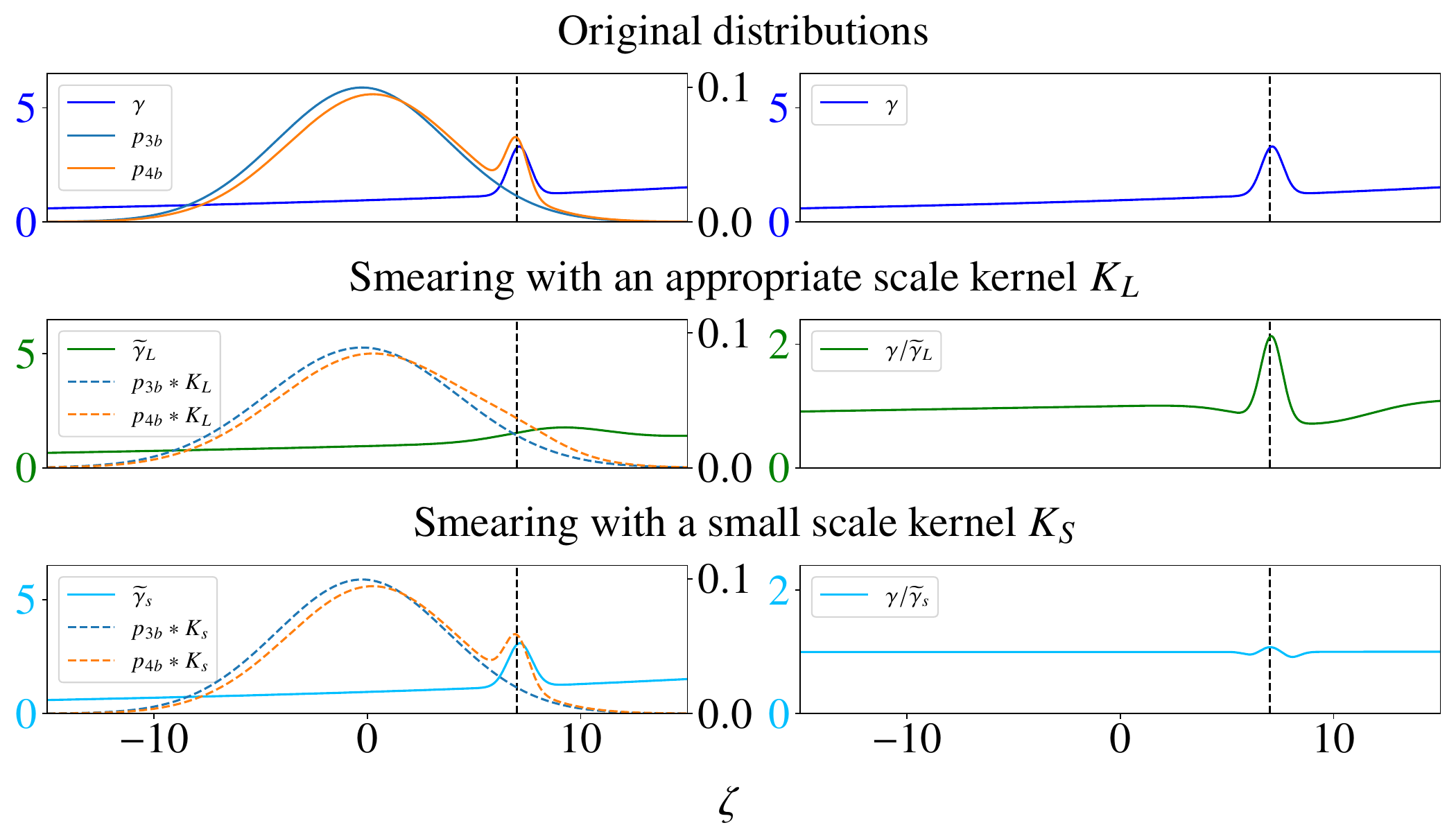}
  \hfill
  \includegraphics[width=0.51\textwidth]{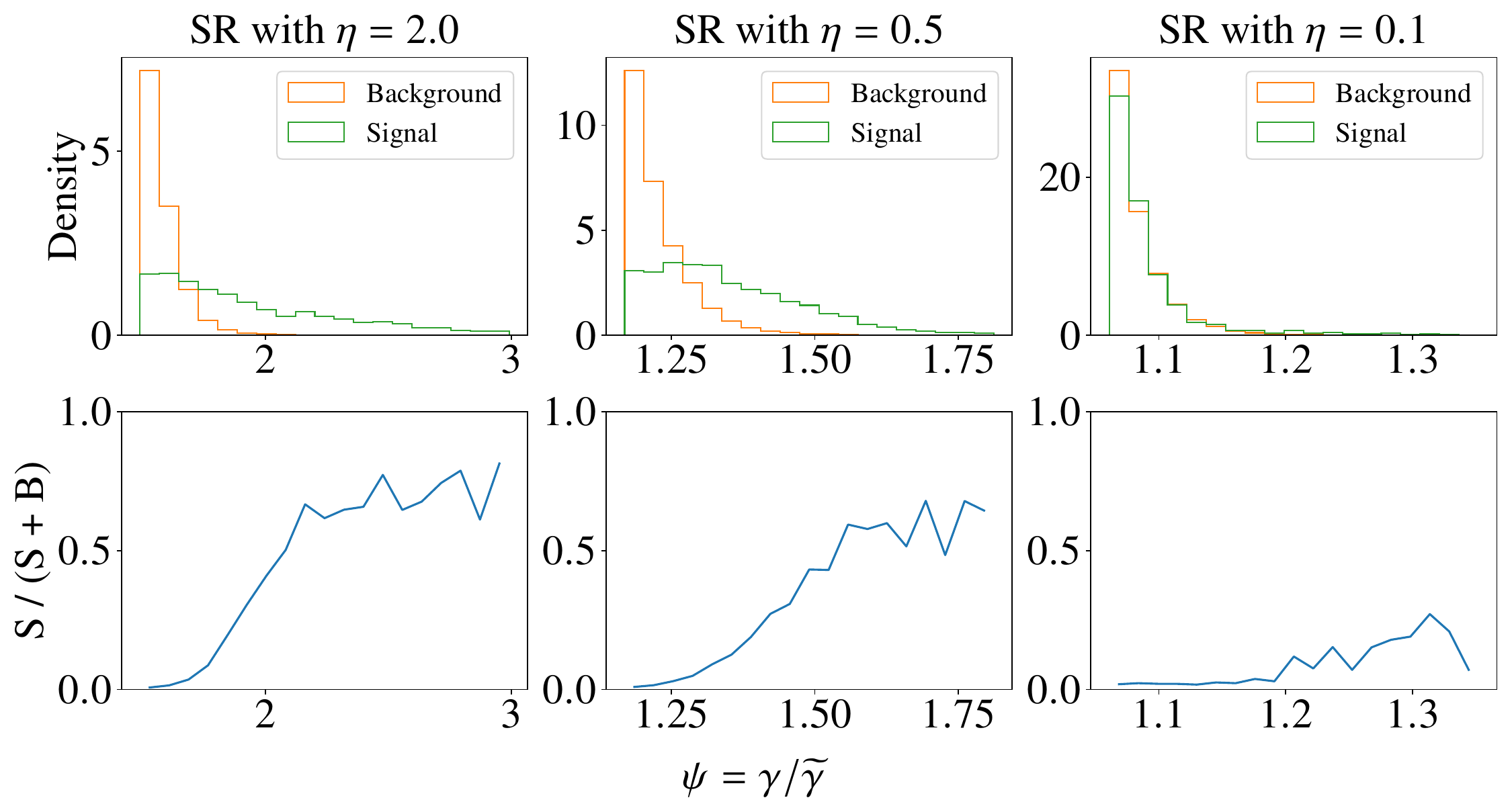}
  \caption{
    A toy example (left panel) shows why the choice of noise scale \(\eta\) should not be too small. 
    When $\eta$ is small, the smeared density ratio \(\gammatilde\) still preserves the high-frequency signal feature (third row, left panel), so thresholding \(\psi = \gamma / \gammatilde\) fails to identify a region rich in signal events.
    Indeed, choosing a smaller \(\eta\) causes signal events to be less concentrated in the tail of \(\psi\) (first row, right panel).
    Consequently, signal events are dominated by background events over the entire range of $\psi$ (second row, right panel), resulting in a loss of power for the hypothesis test.
    }
  \label{fig:smear_toy}
\end{figure}

We take the smoothing kernel to be Gaussian throughout, and write $K_\eta$ for the centered Gaussian density on $\cZ$ with covariance $\eta^2 \diag(\sigma_1^2, \dots, \sigma_d^2)$, where $\sigma_i$ is the standard deviation of the $i$-th representation coordinate and the noise scale $\eta > 0$ sets the bandwidth.
Convolving each density with $K_\eta$ yields a smoothed version of the density ratio that captures only low-frequency components, thereby suppressing features associated with the signal.
We define the convolution of a function $f$ with $K_\eta$ as
\[
    (f * K_\eta)(z) \coloneqq \int_\cZ f(z') K_\eta(z - z') \, dz'
\]
and the ratio of smoothed densities as
\begin{equation}\label{eq:smoothed_density_ratio}
    \gammatilde(x) \coloneqq \frac{(p^\cZ_{4b} * K_\eta)(\zeta(x))}{(p^\cZ_{3b} * K_\eta)(\zeta(x))}
\end{equation}
for \(x \in \cX\), where $p^\cZ_{3b}$ ($p^\cZ_{4b}$) is the probability density function of $3b$ ($4b$) events in the representation space $\cZ$.
To this end, we write 
\begin{equation*}
    \psi(x) = \frac{\gamma(x)}{\gammatilde(x)}.
\end{equation*}
\Cref{fig:smear_toy} illustrates the role of the noise scale: when $\eta$ is too small, $\gammatilde$ retains the high-frequency signal feature, so $\psi$ does not isolate a signal-rich tail.
Both convolved densities in \eqref{eq:smoothed_density_ratio} vanish at the same $\eta^{-d}$ rate as $\eta \to \infty$, and because $p^\cZ_{3b}$ and $p^\cZ_{4b}$ are both probability densities, the leading factors coincide and cancel in the ratio.
Therefore, $\gammatilde$ converges to $1$ and $\psi$ converges to $\gamma$.
We use $\eta = \infty$ to denote this limiting baseline, implemented by setting $\widehat\psi = \widehat\gamma$.

We then define the data-driven SR and CR as
\begin{equation*}
    \cX_s \coloneqq \psi^{-1}([\tau_s, \infty)), \quad
    \cX_c \coloneqq \psi^{-1}([\tau_c, \tau_s)).
\end{equation*}
The thresholds $\tau_s$, $\tau_c$ are specifically designed to control the size of the SR and CR as measured by the proportion of $4b$ events within those regions. 

For evaluating $\psi$, we adopt the ensemble maximum strategy introduced in \Cref{ss:ensemble} to enhance signal concentration in the SR, as shown in the right panel of \Cref{fig:ensemble_of_density_ratio}.
This requires learning $\gammatilde$ multiple times, which can be done efficiently by learning a classifier without directly evaluating the convolution operation.
In particular, we train a classifier to distinguish between the pairs of covariates and labels $(Z_{3b} + \cE_{3b}, 0)$ and $(Z_{4b} + \cE_{4b}, 1)$, where $Z_{3b}$ and $Z_{4b}$ are the representations of $3b$ and $4b$ events, respectively, and $\cE_{3b}, \cE_{4b}$ are i.i.d. noise variables with density $K_\eta$, independent of the representations.
Since the density of a sum of independent random variables is the convolution of their densities, this classifier estimates the ratio of the convolved representation-space densities, namely $\gammatilde$ in \eqref{eq:smoothed_density_ratio}.

\begin{figure}
    \centering
    \includegraphics[width=0.95\textwidth]{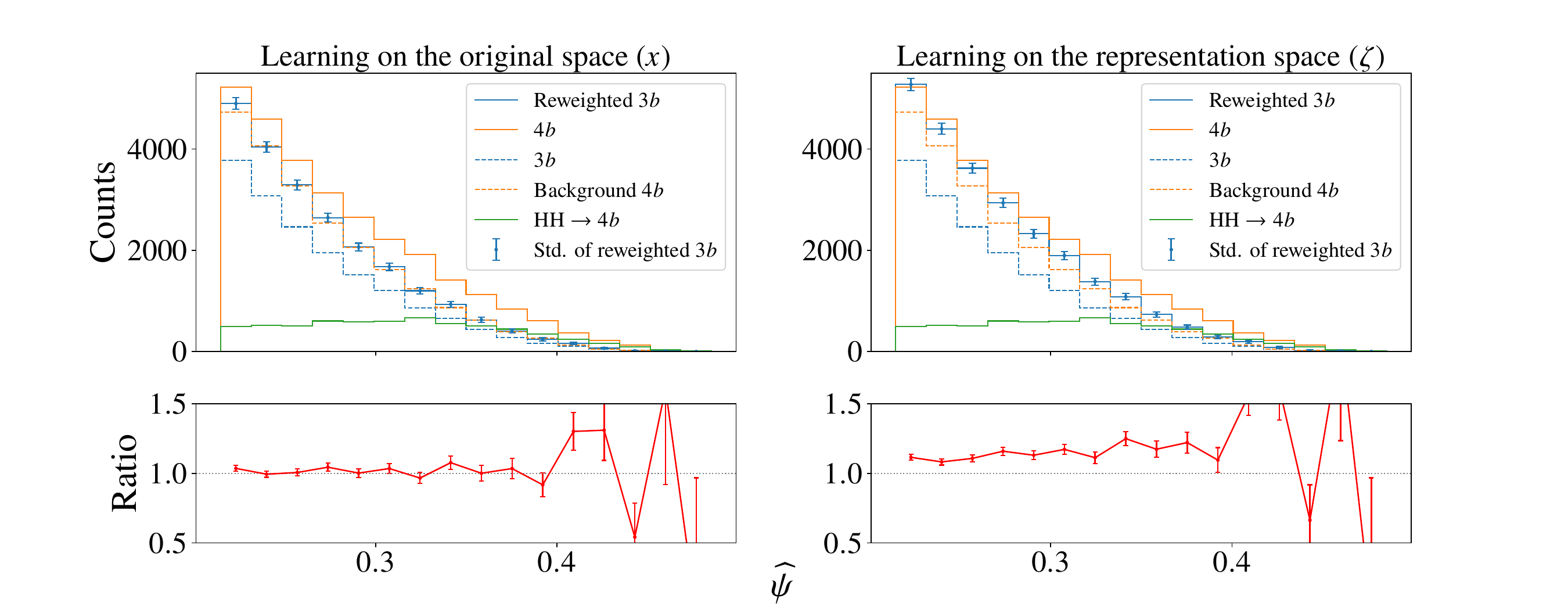}
    \caption{
    Histogram and ratio of estimated background (i.e., reweighted \(3b\)) and true \(4b\) background.
    The choice of space on which to train the density ratio in the CR impacts the estimated background distribution (i.e., counts of reweighted $3b$ events in each bin) for $4b$ events. 
    Learning the density ratio on the original space (left) provides a reasonable estimate of the background $4b$ distribution.
    Training the density ratio on the representation space (right), on the other hand, introduces a strong bias toward higher values in the presence of signal events. 
    }
    \label{fig:on_which_to_learn}
\end{figure}

\begin{figure}
    \centering
    \includegraphics[width=0.6\textwidth]{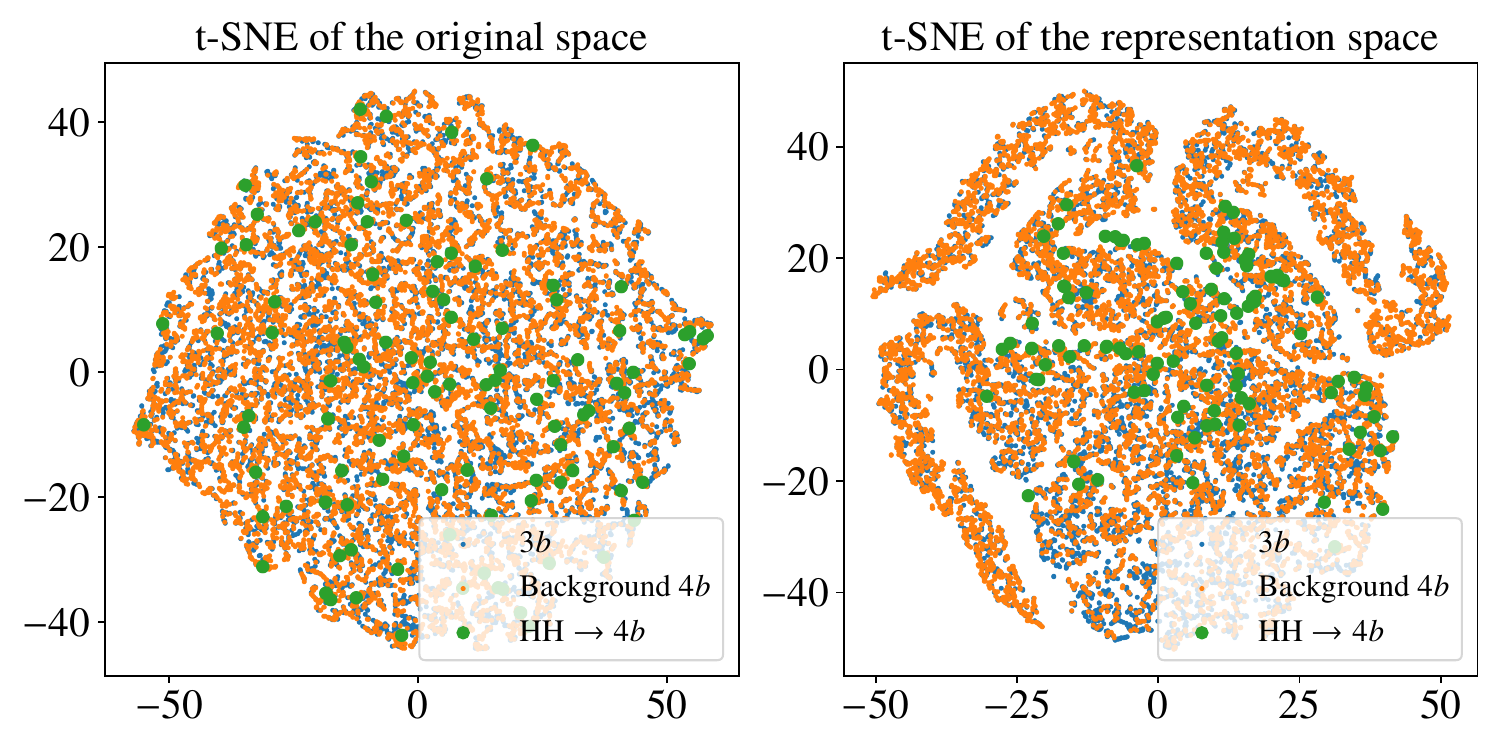}
    \caption{
      t-SNE plots of original features (left) and representations (right) in the CR with colors based on types of events.
      Signal events in the representation space are more concentrated in a specific area compared to the original space, leading to a larger bias in the background estimator when it is learned based on the representations.
    }
    \label{fig:tsne_original_repr}
\end{figure}

We employ the same classifier architecture used in \Cref{ss:learning_density_ratio} to learn $\widehat{\gamma}_{\text{CR}}$, training the classifier only on CR events in the original space of $x$.
It is worth noting that, although it is possible to train a classifier to learn the ratio based on the representations $\zeta$, we opt not to do so. As illustrated in \Cref{fig:on_which_to_learn}, contamination of the CR by signal events leads to a biased estimate of the background $4b$ distribution when training is performed on the representations, whereas training on the original space does not suffer from this issue.
One possible explanation for this discrepancy is that signal events are concentrated in a localized region of the representation space, while they are spread out in the original space, as depicted in \Cref{fig:tsne_original_repr}.
The localization that makes smearing effective in the representation space therefore also makes a background estimator learned there more vulnerable to signal contamination, which is why we smear in the representation space but learn $\widehat{\gamma}_{\text{CR}}$ in the original one.
Consequently, contamination of the background estimator by signal events is amplified in the representation space, leading to larger bias.

As a final remark, we note that the dataset is partitioned into two splits (two vertical splits in \Cref{fig:workflow}): the first is used to learn $\widehat\psi$ and define the SR and CR, while the second is used to learn $\widehat{\gamma}_{\text{CR}}$ and perform hypothesis tests.
Consequently, the SR can be treated as fixed in the subsequent analysis, and events within the SR retain the Poisson structure of the underlying event process.
Precise procedures for learning the ratio of smoothed densities, evaluating $\psi$ with an ensemble maximum, and selecting two thresholds to define the SR and CR are outlined in \Cref{alg:define_regions}.

\begin{algorithm}[t]
    \caption{Defining the SR and CR}
    \begin{algorithmic}[1]
        \Procedure{LearnSmoothedDensityRatio}{$\bX$, $\bY$, $\zeta$, $\eta$}
        \State Compute standard deviation $\sigma_i$ of $\{\zeta_i(x)\,|\,x \in \bX\}$ for $i \in \{1, \dots, d\}$
        \State Sample random noise $\cE_x \stackrel{\text{i.i.d.}}{\sim} \cN(0, \eta^2 \diag(\sigma_1^2, \dots, \sigma_d^2))$ for each $x \in \bX$
        \State $\widehat{\widetilde{\gamma}} \gets \textsc{LearnDensityRatio}(\{\zeta(x) + \cE_x\,|\,x \in \bX \}, \bY)$
        \State \Return $\widehat{\widetilde{\gamma}}$
        \EndProcedure
        \Procedure{MaxEnsembleCriterion}{$\bX$, $\bY$, $\eta$, $E$}
        \For {different random seeds $i = 1, \dots, E$} \Comment{Ensemble for defining the SR}
        \State $\widehat{\gamma}^i$, $\zeta^i \gets \textsc{LearnDensityRatioAndRepresentation}(\bX, \bY)$
        \State $\hatgammatilde^i \gets \textsc{LearnSmoothedDensityRatio}(\bX, \bY, \zeta^i, \eta)$
        \State $\widehat{\psi}^i \gets (\widehat{\gamma}^i \circ \zeta^i) / (\hatgammatilde^i \circ \zeta^i)$
        \EndFor
        \State \Return $\max_{i=1}^E \widehat{\psi}^i$ \Comment{Ensemble maximum}
        \EndProcedure
        \Procedure{DefineRegions}{$\bX$, $\widehat{\psi}$, $s_{\text{SR}}$, $s_{\text{CR}}$}
        \State $\bX_{4b} \gets \{x_{4b}\,|\,x_{4b} \in \bX \text{ and } x_{4b} \text{ is a } 4b \text{ event}\}$, $N_{4b} \gets \abs{\bX_{4b}}$
        \State $\tau_s \gets$ empirical $(1 - s_{\text{SR}})$-quantile of $\{\widehat{\psi}(x_{4b})\,|\,x_{4b} \in \bX_{4b}\}$
        \State $\tau_c \gets$ empirical $(1 - s_{\text{SR}} - s_{\text{CR}})$-quantile of $\{\widehat{\psi}(x_{4b})\,|\,x_{4b} \in \bX_{4b}\}$
        \State \Return $\cX_s \coloneqq \widehat{\psi}^{-1}([\tau_s, \infty)), \quad \cX_c \coloneqq \widehat{\psi}^{-1}([\tau_c, \tau_s))$
        \EndProcedure
    \end{algorithmic}
    \label{alg:define_regions}
\end{algorithm}

\subsection{Hypothesis Testing with Systematic Uncertainty}
\label{ss:hypothesis_testing}

Having defined the SR and CR and established a way to estimate the background distribution within the SR, the remaining ingredient is a test that detects the presence of a signal in the SR.
A common approach to hypothesis testing in high-energy physics involves binned counting tests based on the Poisson-process assumption for collision events \citep{ATLASCollaboration2010_search,ATLAS2012_observation,Cowan2011_likelihood,CMS2012_observation}.
However, this type of test introduces an additional hyperparameter, the number of bins, to which the test can be sensitive.
To avoid this issue, we opt for Kolmogorov--Smirnov (KS) type test statistics, which do not rely on additional binning hyperparameters \citep{Massey1951_kolmogorov}.

To construct the test, we compare the empirical cumulative distribution function (CDF) of $\widehat\psi$ for $4b$ events in the SR with the CDF estimated from $3b$ events in the SR reweighted by $\widehat\gamma_{\mathrm{CR}}$.
We formulate two empirical CDFs as
\begin{equation}
\begin{aligned}
    \widehat F_{3b}(y)
    &=\frac{\sum_{X_{3b}^i\in\cX_s}\widehat\gamma_{\mathrm{CR}}(X_{3b}^i)
    \mathbf1\{\widehat\psi(X_{3b}^i)\leq y\}}
    {\sum_{X_{3b}^i\in\cX_s}\widehat\gamma_{\mathrm{CR}}(X_{3b}^i)},\\
    \widehat F_{4b}(y)
    &=\frac{\sum_{X_{4b}^j\in\cX_s}\mathbf1\{\widehat\psi(X_{4b}^j)\leq y\}}
    {|\{j\,|\,X_{4b}^j\in\cX_s\}|}.
\end{aligned}
\label{eq:weighted_empirical_cdf}
\end{equation}
Here, $\cX_s$ is the SR and $\mathbf1$ is the indicator function; physical event weights, suppressed in the CDF formulas here and below, are incorporated as described in \Cref{s:weights}.
The test statistic is the maximum difference between the empirical CDFs:
\begin{equation}
    T_0=\max_{y\in\R}\left|\widehat F_{3b}(y)-\widehat F_{4b}(y)\right|.
    \label{eq:ks_observed_statistic}
\end{equation}

However, imperfect classifier training may cause $\widehat\gamma_{\mathrm{CR}}$ to deviate from the true background density ratio in the SR.
We therefore introduce a nuisance parameter to account for residual systematic mismodeling in the hypothesis test \citep{Conway2011_nuisance}.
We posit that the density ratio can be adjusted by a simple multiplicative correction represented as an affine function of $\log\widehat\psi$:
\begin{equation}
\begin{aligned}
    \widehat\gamma_{\mathrm{CR}}^\delta(x)
    &=\bigl[1+\delta\{\log\widehat\psi(x)-A\}\bigr]\widehat\gamma_{\mathrm{CR}}(x),\\
    A &=
    \frac{\sum_{X_{3b}^i\in\cX_s} (\widehat\gamma_{\mathrm{CR}}w_{3b})(X_{3b}^i)\log\widehat\psi(X_{3b}^i)}{\sum_{X_{3b}^i\in\cX_s} (\widehat\gamma_{\mathrm{CR}}w_{3b})(X_{3b}^i)},
\end{aligned}
    \label{eq:affine_correction}
\end{equation}
for some $\delta\in\R$.
Here $A$ is the weighted mean of $\log\widehat\psi$ for the $3b$ events in the SR with weights before correction: centering at $A$ preserves the original total $3b$ weight for every $\delta$.
We restrict $\delta$ to the set $\Delta$ for which the affine factor is nonnegative over the observed $3b$ score range.

% For the population hypotheses, we consider corrections whose affine factors are nonnegative throughout the SR.
For such corrections, let $F_{3b}^\delta$ and $F_{4b}$ denote the population CDFs of $\widehat\psi$ in the SR under the background distribution corrected by $\delta$ and the observed $4b$ distribution, respectively.
We test the following composite distributional hypotheses:
\begin{equation}
\begin{aligned}
    H_0^{\mathrm{comp}} &: F_{3b}^\delta=F_{4b}
    \quad\text{for some such correction }\delta,\\
    H_1^{\mathrm{comp}} &: F_{3b}^\delta\neq F_{4b}
    \quad\text{for every such correction }\delta.
\end{aligned}
\label{eq:composite-null}
\end{equation}
In particular, we compute the empirical CDF of $3b$ events with corrected weights
\begin{align*}
    \widehat F_{3b}^\delta(y)
    =\frac{\sum_{X_{3b}^i\in\cX_s}\widehat\gamma_{\mathrm{CR}}^\delta(X_{3b}^i)
    \mathbf1\{\widehat\psi(X_{3b}^i)\leq y\}}
    {\sum_{X_{3b}^i\in\cX_s}\widehat\gamma_{\mathrm{CR}}^\delta(X_{3b}^i)}
\end{align*}
for every admissible correction $\delta$, and compute the KS test statistic under this correction:
\begin{align*}
    T_0(\delta)
    = \max_{y \in \R}
    \left| \widehat F_{3b}^\delta(y) - \widehat F_{4b}(y) \right|.
\end{align*}

The null distributions of the test statistics after correction are approximated by Poisson resampling from the pooled score--weight pairs in the SR.
Each replicate $b$ produces two resampled samples whose total weights match the observed $3b$ and $4b$ weight totals in expectation.
$T_b(\delta)$ is the KS-type statistic comparing the weighted empirical CDFs of the two resampled samples, with tied scores handled as specified in \Cref{s:continuous_affine_bootstrap}.
The $p$-value with correction $\delta$ is
\begin{equation}
    p(\delta)=\frac{1+\sum_{b=1}^B\mathbf1\{T_b(\delta)\geq T_0(\delta)\}}{B+1},
    \label{eq:affine_bootstrap_ks_main}
\end{equation}
where we use the same $B$ sets of Poisson counts for every admissible correction.
We obtain the composite $p$-value by maximizing over all admissible corrections:
\begin{equation}
    p=\sup_{\delta \in \Delta} p(\delta).
    \label{eq:affine_nuisance_pvalue}
\end{equation}
We reject when $p\leq\alpha$: rejection requires rejecting the null hypothesis for every admissible affine correction.
This composite-null construction controls the Type I error rate at level $\alpha$ if the maximization includes a correction under which the null hypothesis holds and the bootstrap test is valid for that correction \citep{BergerBoos1994_pvalues}.
The bootstrap and maximization procedures are detailed in \Cref{s:continuous_affine_bootstrap}; combining region construction, background estimation, and this composite KS test gives the complete procedure in \Cref{alg:complete_hypothesis_testing}.

% We retain the original class-weight totals when pooling; each Poisson rate is the corresponding class total divided by the pooled total.
% We repeat this process $B$ times to obtain bootstrap statistics $T_1,\ldots,T_B$.
% The precise resampling and cumulative-step conventions are given in \Cref{s:continuous_affine_bootstrap}.

% \subsection{Systematic Error Correction}
% \label{ss:systematic_error_correction}

% For each admissible $\delta$, replacing $\widehat\gamma_{\mathrm{CR}}$ by $\widehat\gamma_{\mathrm{CR}}^\delta$ in \eqref{eq:weighted_empirical_cdf} gives the corrected CDF $\widehat F_{3b}^\delta$.
% Under the corrected no-signal null, there is an admissible $\delta_\star$ such that $F_{3b}^{\delta_\star}=F_{4b}$, where $F_{3b}^{\delta}$ and $F_{4b}$ denote the population score CDFs for the corrected background and the $4b$ sample, respectively.
% At this correction, $\widehat F_{3b}^{\delta_\star}$ and $\widehat F_{4b}$ estimate the same population CDF.

\begin{algorithm}[t]
\caption{Complete algorithm for hypothesis testing}
\begin{algorithmic}[1]
    \State \textbf{Input: } Event samples $\bX$ and corresponding labels $\bY$, number of ensemble members $E$, noise scale $\eta$, size of SR $s_{\text{SR}}$, size of CR $s_{\text{CR}}$, number of bootstrap replicates $B$
    \State \textbf{Output: } $p$-value for testing the composite null hypothesis $H_0^{\mathrm{comp}}$ in \eqref{eq:composite-null}
    \State Partition $\bX$ into disjoint subsets $\bX_1$ and $\bX_2$, and partition $\bY$ accordingly
    \State $\widehat{\psi} \gets \textsc{MaxEnsembleCriterion}(\bX_1, \bY_1, \eta, E)$
    \State $\cX_s$, $\cX_c \gets \textsc{DefineRegions}(\bX_1, \widehat{\psi}, s_{\text{SR}}, s_{\text{CR}})$
    \State $\bX_{\text{CR}} \gets \bX_2 \cap \cX_c$, $\bX_{\text{SR}} \gets \bX_2 \cap \cX_s$
    \State $\widehat{\gamma}_{\text{CR}} \gets \textsc{LearnDensityRatio}(\bX_{\text{CR}}, \bY_{\text{CR}})$
    \State Compute $p$ from $\bX_{\text{SR}}$ using pooled Poisson resampling and composite-null maximization in \eqref{eq:affine_bootstrap_ks_main} and \eqref{eq:affine_nuisance_pvalue}
    \State \Return $p$
\end{algorithmic}
\label{alg:complete_hypothesis_testing}
\end{algorithm}

\section{Results}
\label{s:results}

We present our experimental results in two stages. In the first two subsections, we evaluate the proposed method on the non-resonant $\mathrm{HH} \to 4b$ process, our primary signal benchmark. \Cref{ss:res_concentration_of_signal_events_in_the_SR} quantifies signal concentration in the SR and \Cref{ss:res_hypothesis_testing} evaluates hypothesis testing performance, with a focus on Type I error control and statistical power.

In \Cref{ss:res_generalization}, we test the versatility of the method by applying it to resonant di-Higgs production $X \to \mathrm{HH} \to 4b$ with $m_X = 400\,\mathrm{GeV}$ and non-resonant \(\mathrm{ZH} \to 4b\) without re-tuning the parameters. 
These benchmarks have different signal kinematic properties from the non-resonant \(\mathrm{HH} \to 4b\) case, allowing us to confirm that the method generalizes across distinct types of signals while preserving statistical validity.

\subsection{Concentration of signal events in the SR}
\label{ss:res_concentration_of_signal_events_in_the_SR}

In this section, we demonstrate that our method successfully achieves concentration of signal events in the SR. 
Similar to \Cref{fig:ensemble_of_density_ratio}, we examine a plot of $S(\cX_s)$ against $P_{4b}(\cX_s)$ to evaluate signal concentration. The goal is to attain a high proportion of signal events (i.e., $S(\cX_s)$) for a given SR size (i.e., $P_{4b}(\cX_s)$). In an ideal case, the curve should rapidly increase to $1$ as the SR size increases.

We conduct experiments with $10^6$ $3b$ samples and a similar number of $4b$ events with the total weight the same for both $3b$ and $4b$ events. 
We use half of all the samples as $\bX_1$ to estimate $\gamma$ and $\gammatilde$, and leave the remaining samples for hypothesis testing.
Each dataset is assembled from three separate pools of simulated events: $2{,}013{,}103$ $3b$ events, $2{,}202{,}261$ background $4b$ events, and a pool of signal events for the process under study.
We repeat the experiment $100$ times, using a new random seed each time to draw a fresh subsample without replacement from each pool: the $3b$ events are drawn to the specified count, while the background $4b$ and signal events are taken in random order until their target total weights are reached, the latter set by the designated signal ratio $\epsilon$.

\begin{figure}[t]
  \centering
  \includegraphics[width=0.4\textwidth]{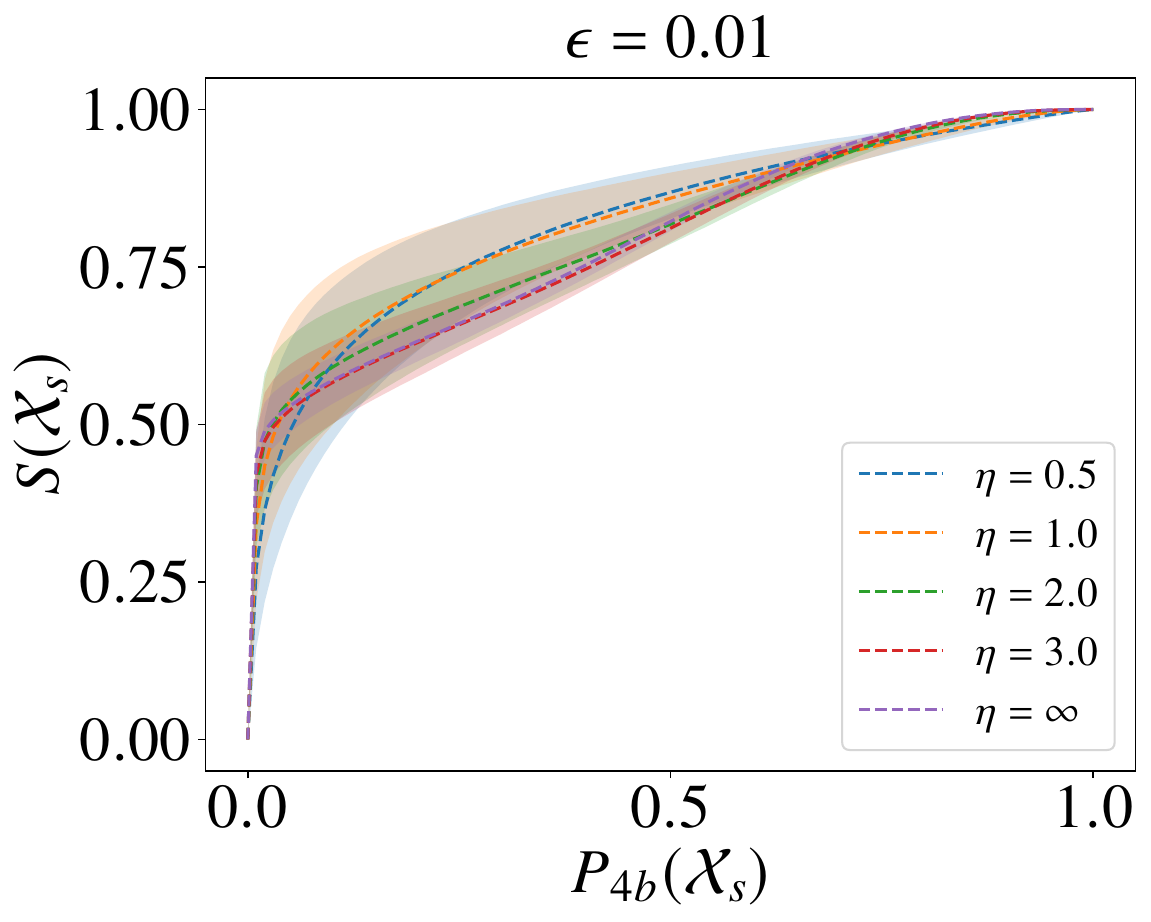}
  \includegraphics[width=0.4\textwidth]{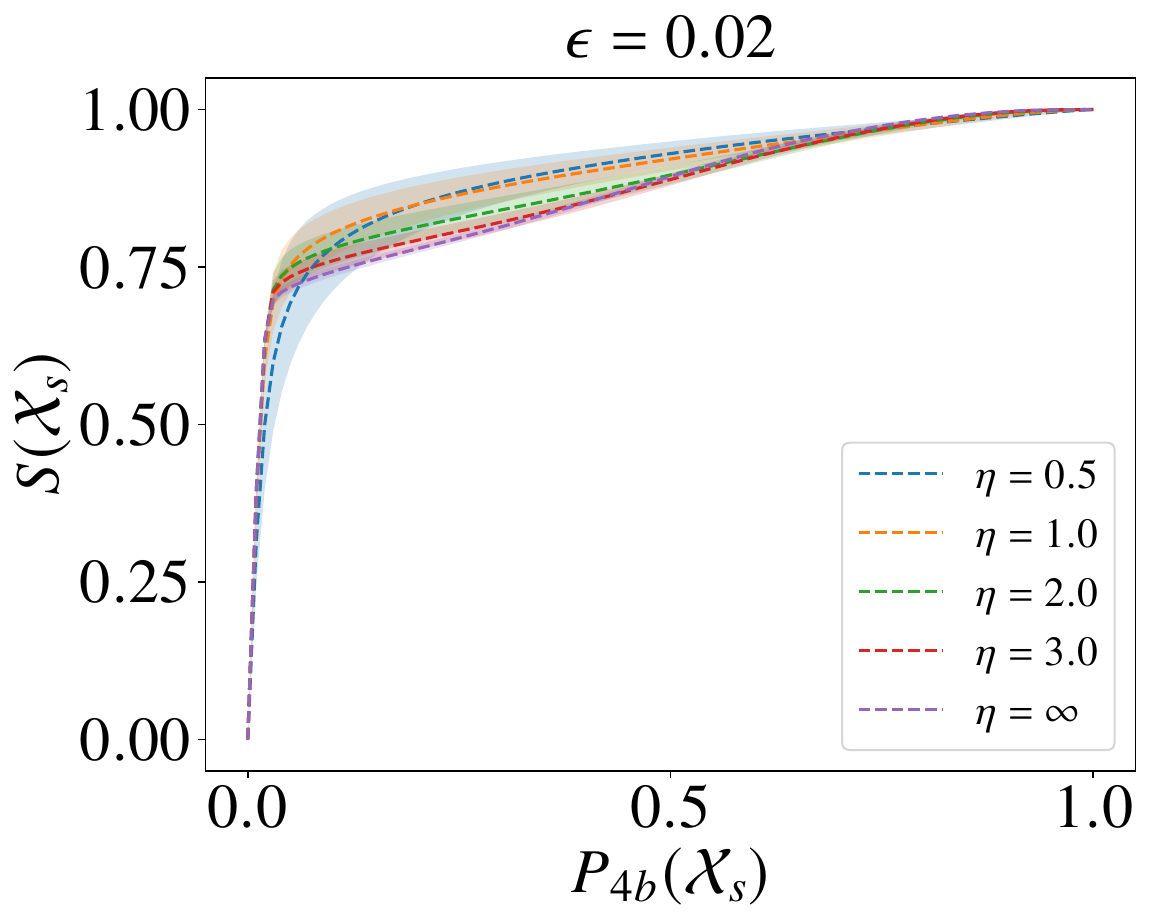}
  \caption{The SR efficiency of the non-resonant $\mathrm{HH} \to 4b$ signal with signal ratio $\epsilon = 0.01$ (left) and $\epsilon = 0.02$ (right) for different noise scales $\eta$.
  Dashed lines represent the average measured across $100$ repeated experiments, and the shaded regions represent the corresponding standard deviations.
  The noise scales give comparable concentration overall.
  On the left side of the curves, where $P_{4b}(\cX_s)$ is small and the SR consists of the highest-$\psi$ events, $\eta = 0.5$ shows the least concentration and overtakes the larger scales only as the SR widens.}
  \label{fig:SR_efficiency_non_resonant_noise_scale}
\end{figure}

\Cref{fig:SR_efficiency_non_resonant_noise_scale} shows curves of $S(\cX_s)$ against $P_{4b}(\cX_s)$ for different values of the noise scale $\eta$ at two signal ratios $\epsilon \in \{0.01, 0.02\}$.
The test gains power when signal concentrates in the tail of $\psi$ within the SR (\Cref{fig:smear_toy}).
The left end of these curves, where $P_{4b}(\cX_s)$ is small, reflects that concentration, since it is set by the highest-$\psi$ events.
At $\epsilon = 0.01$, $\eta = 0.5$ is the lowest of the noise scales at $s_{\text{SR}} = 0.05$ and overtakes $\eta = 2.0$ and $\eta = 3.0$ only as the SR widens, so it concentrates signal least in the tail.
We therefore rule out the smallest noise scales and use $\eta = 2.0$ in the hypothesis tests below, reporting the other scales in \Cref{tab:hypothesis_testing_noise_scale}.

\begin{figure}
    \centering
    \includegraphics[width=0.4\textwidth]{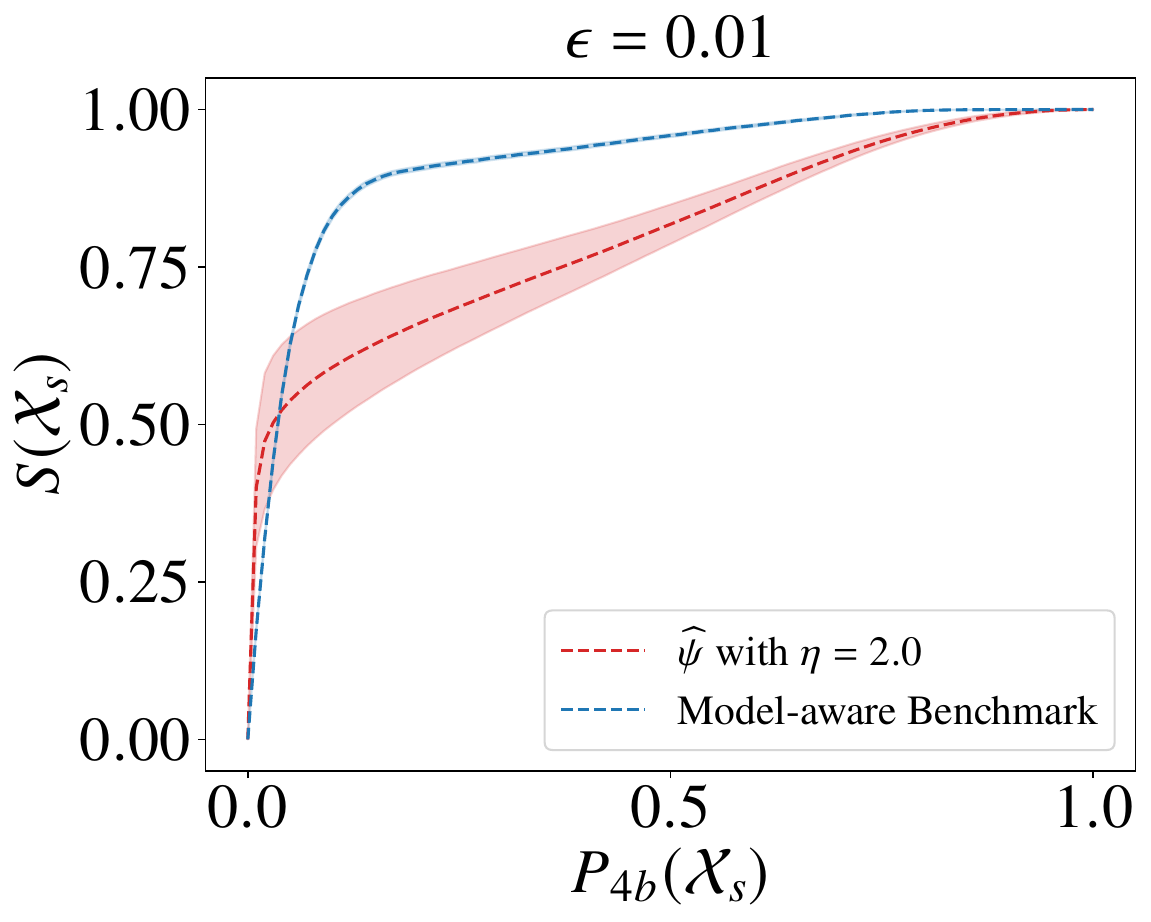}
    \includegraphics[width=0.4\textwidth]{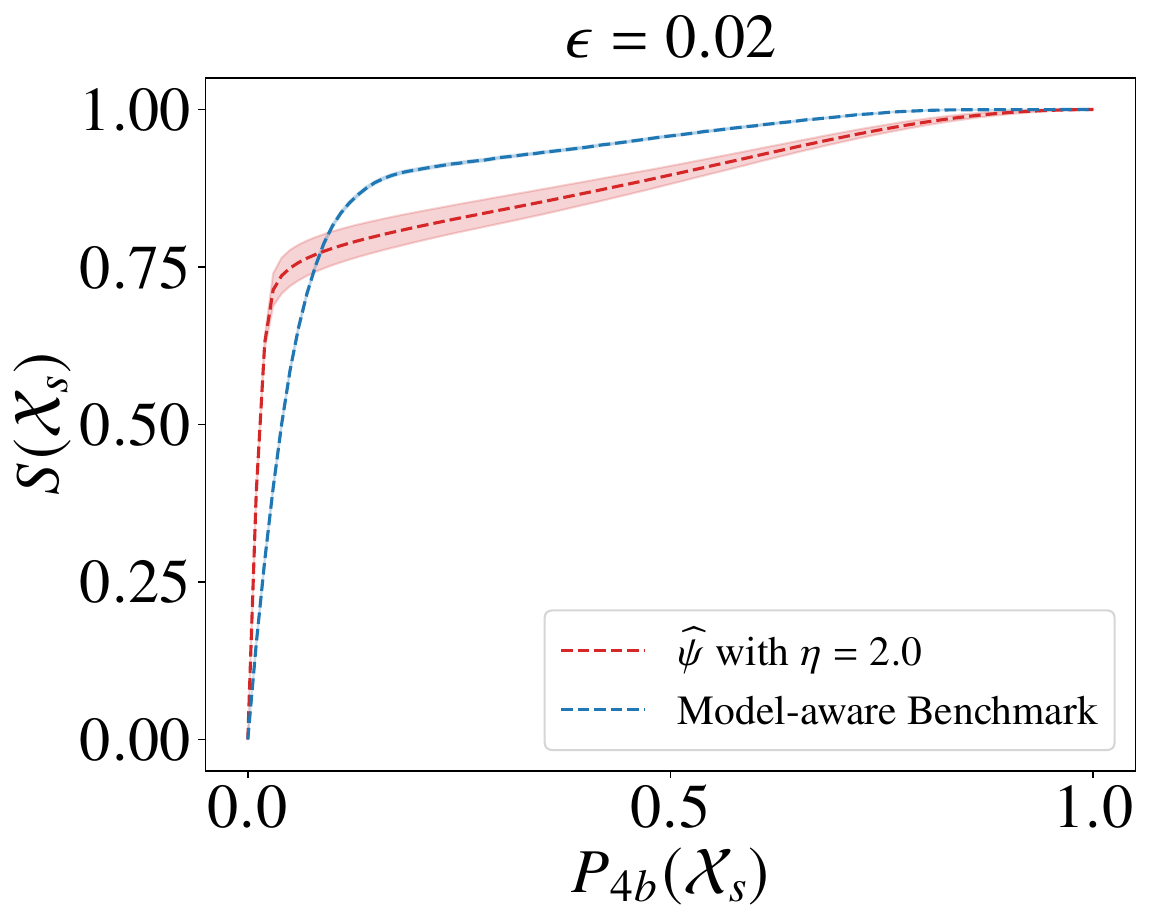}
    \caption{Comparison with the baseline SR proposed by \citet{Bryant2018_search}.
    Dashed lines represent the average measured across $100$ repeated experiments, and the shaded region represents the standard deviation of our method.
    While the baseline, which has access to a priori knowledge of the location of the signal, shows better performance when $P_{4b}(\cX_s)$ is large, our method matches or exceeds it once $P_{4b}(\cX_s)$ is small.}
    \label{fig:SR_comparison}
\end{figure}

With this choice, \Cref{fig:SR_comparison} compares our method with a model-aware benchmark SR proposed by \citet{Bryant2018_search}.
The baseline captures events whose two intermediate particles both have invariant masses close to that of the Higgs boson, which requires knowledge of the physical properties of the signal.
Our method is model-agnostic and needs no such knowledge, yet reaches competitive signal concentration and matches or exceeds the baseline when the SR is small.
That is the regime that matters for the test, since concentration in the tail of $\psi$ is what gives it power.

% \begin{figure}
%     \centering
%     \includegraphics[width=0.9\textwidth]{figures/SR_efficiency.pdf}
%     \caption{\memo{This figure should be replaced.} Concentration of signal events in the SR for the $\mathrm{HH} \to 4b$ data. $x$ axis represents the proportion of $4b$ events in the SR (measured by $P_{4b}(\cX_s)$) and $y$ axis represents the proportion of signal events in the SR (measured by $S(\cX_s)$). 
%     As defined previously, ${n_{3b}}$, $\epsilon$, and $\eta$ are the number of $3b$ events, the signal ratio, and the scale of the convolution kernel, respectively.
%     Bold lines and shaded areas represent (mean) $\pm$ (standard deviation) measured through $10$ repeated experiments.
%     }
%     \label{fig:SR_efficiency}
% \end{figure}

\subsection{Hypothesis Testing Performance}
\label{ss:res_hypothesis_testing}
In this section, we provide results that demonstrate the validity and power of the hypothesis test introduced in \Cref{ss:hypothesis_testing} with the data-driven SR set following our method. 
With the same setup as in the previous subsection and the noise scale $\eta = 2.0$, we report its Type I error rate ($\epsilon = 0$) and power for different signal ratios $\epsilon$ varying across $\{0.005, 0.0075, 0.01, 0.02\}$. 
We set $s_{\text{CR}} = 1 - s_{\text{SR}}$, with $s_{\text{SR}}$ taking values from $\{0.05, 0.1, 0.15, 0.2\}$. 
At significance level $\alpha = 0.05$, we observe rejection rates across $100$ repeated experiments for each setup with $B = 1000$ bootstrap replicates, with the repetitions drawn as described in \Cref{ss:res_concentration_of_signal_events_in_the_SR}.

\begin{figure}
  \centering
  \includegraphics[width=0.7\textwidth]{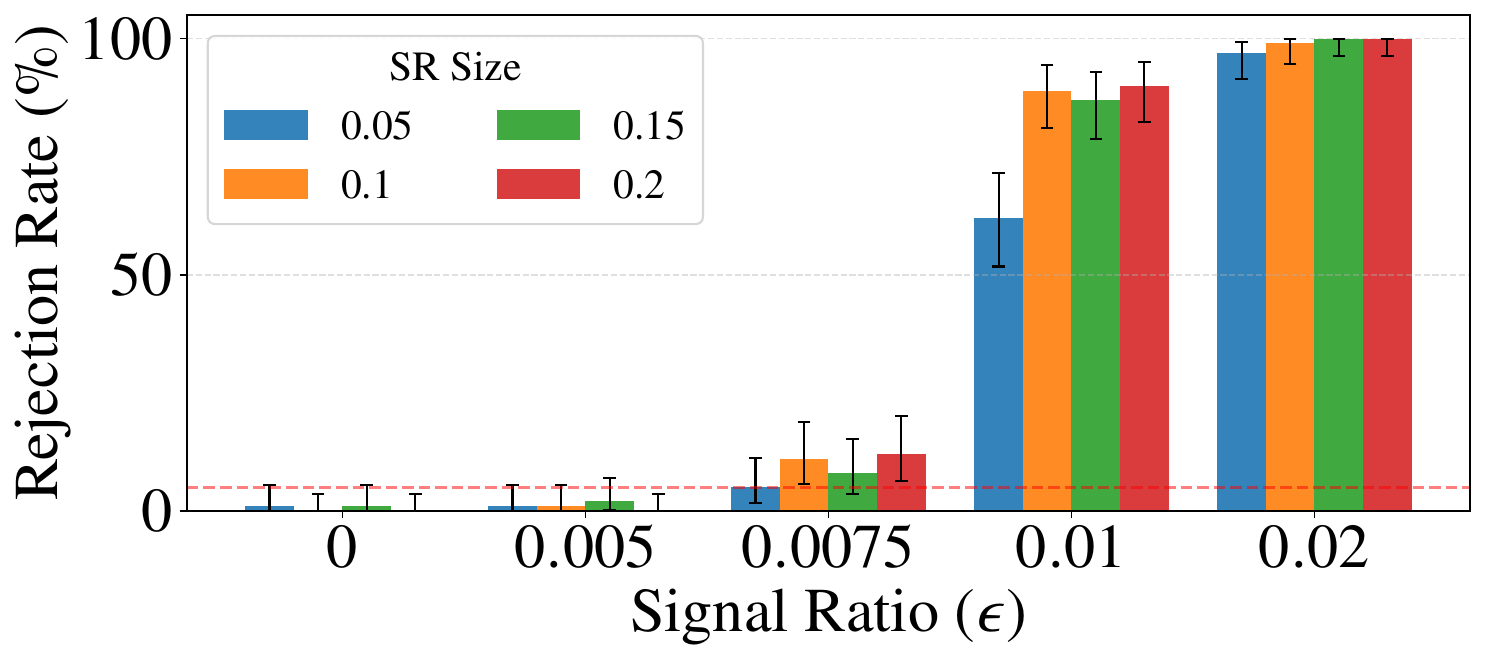}
  \caption{Percentage of experiments rejecting the null hypothesis of \eqref{eq:sr_ratio_test} with non-resonant $\mathrm{HH} \to 4b$ signal for the hypothesis test at significance level $\alpha = 0.05$ with noise scale $\eta = 2.0$.
    Error bars are Clopper--Pearson $95\%$ intervals based on the $100$ repetitions. $\epsilon = 0$ corresponds to the null case, for which the observed rejection rates are below the nominal level $\alpha = 0.05$.}
    \label{fig:hypothesis_testing}
\end{figure}

Under the null hypothesis ($\epsilon = 0$), the empirical rejection rates for the four SR sizes $s_{\text{SR}} \in \{0.05, 0.1, 0.15, 0.2\}$ in \Cref{fig:hypothesis_testing} are $0.01$, $0.00$, $0.01$, and $0.00$, respectively.
Across the four noise scales and four SR sizes in \Cref{tab:hypothesis_testing_noise_scale}, only $7$ of the $1600$ null experiments reject ($0.44\%$), indicating conservative behavior at the nominal level $\alpha = 0.05$ over the settings considered.
Power becomes substantial at $\epsilon=0.01$ and is close to one at $\epsilon=0.02$.
The other noise scales in \Cref{tab:hypothesis_testing_noise_scale} follow the ordering anticipated in \Cref{ss:res_concentration_of_signal_events_in_the_SR}: at $\epsilon = 0.01$, $\eta = 0.5$ rejects in $0.42$ to $0.52$ of the experiments, against $0.72$ to $0.97$ for $\eta = 3.0$.

% \begin{table}[t]
%     \centering
%     \begin{tabular}{c|c|c|c|c|c}
%         \toprule
%         Ratio of $4b$ in SR & $\epsilon = 0$ & $\epsilon = 0.005$ & $\epsilon = 0.0075$ & $\epsilon = 0.01$ & $\epsilon = 0.02$ \\
%         \midrule
%         $0.05$ & $0.02$ & $0.05$ & $0.17$ & $0.89$ & $1.00$ \\
%         $0.1$ & $0.02$ & $0.03$ & $0.28$ & $0.94$ & $1.00$ \\
%         $0.15$ & $0.02$ & $0.04$ & $0.24$ & $0.94$ & $1.00$ \\
%         $0.2$ & $0.02$ & $0.02$ & $0.27$ & $0.95$ & $1.00$ \\
%         \bottomrule
%     \end{tabular}
%     \vspace{0.5em}
%     \caption{Percentage of rejecting the null hypothesis of \eqref{eq:sr_ratio_test} with non-resonant $\mathrm{HH} \to 4b$ signal for the hypothesis testing at significance level $\alpha = 0.05$.
%     The noise scale used in learning the smoothed density ratio $\gammatilde$ is set to $\eta = 2.0$ and $\epsilon$ is the signal ratio.
%     $\epsilon = 0$ corresponds to the null case, which exhibits successful Type I error control below the desired level. }
%     \label{tab:hypothesis_testing}
% \end{table}

\subsection{Generalization to Other Types of Signal Events}
\label{ss:res_generalization}

\begin{figure}
  \centering
  \includegraphics[width=0.4\textwidth]{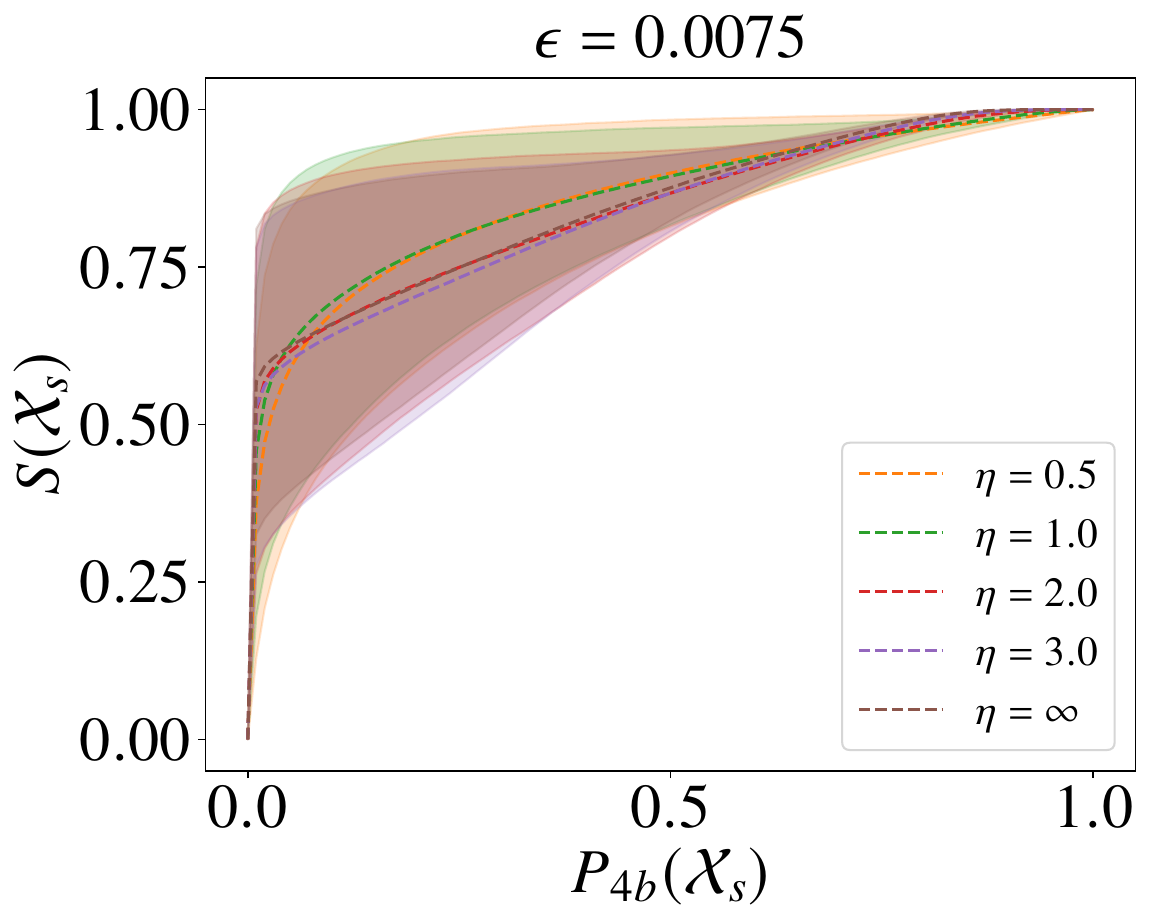}
  \includegraphics[width=0.4\textwidth]{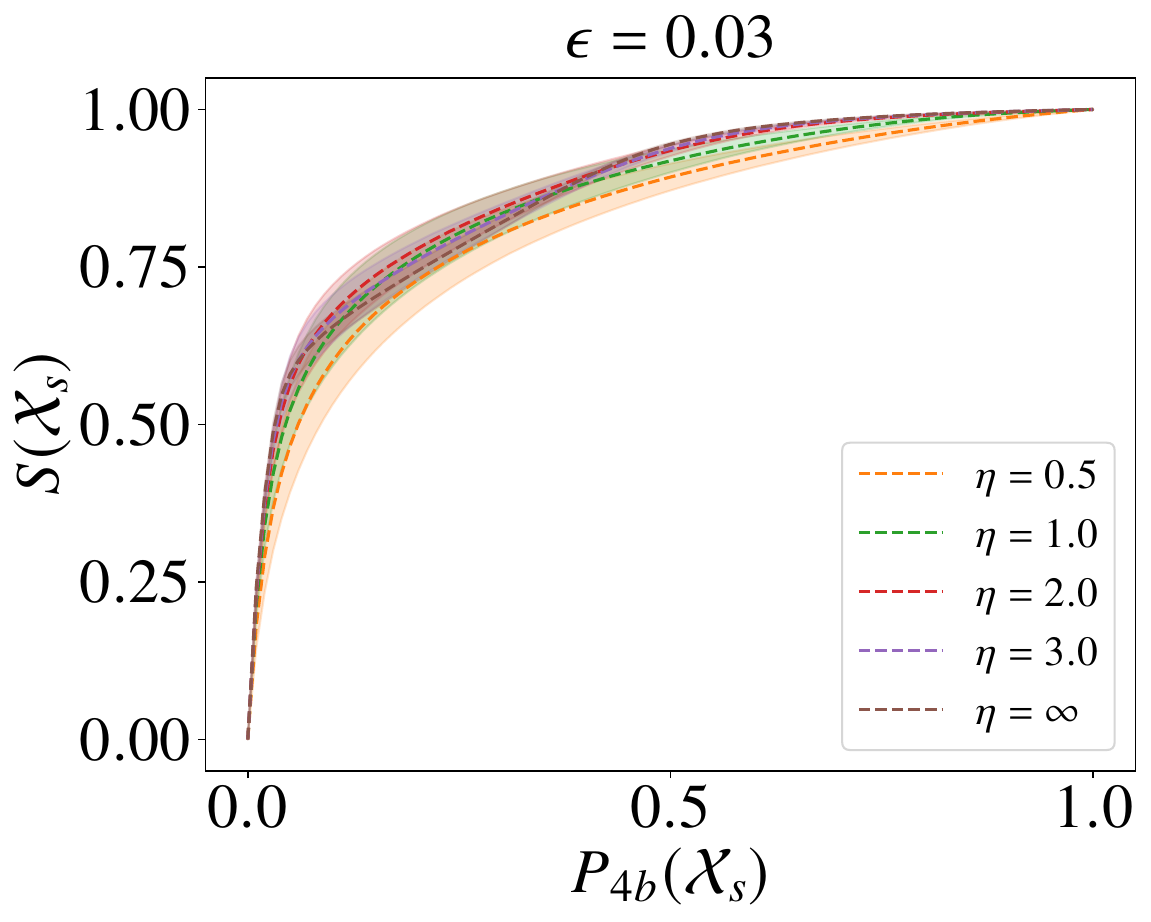}
  \caption{The SR efficiency of the resonant $X \to \mathrm{HH} \to 4b$ signal at $400\,\mathrm{GeV}$ (left) and non-resonant $\mathrm{ZH} \to 4b$ (right) with signal ratios $\epsilon = 0.0075$ and $\epsilon = 0.03$, respectively.
  Dashed lines represent the average measured across $100$ repeated experiments, and the shaded regions represent the corresponding standard deviations.
  The plot demonstrates the adaptivity of our method to different types of signals and its robustness to the choice of the noise scale.
  }
  \label{fig:SR_efficiency_others}
\end{figure}

We evaluate the versatility of our method by investigating its effectiveness in maintaining the concentration of signal events when applied to different types of signal processes. Specifically, we consider two representative cases: resonant $X \to \mathrm{HH} \to 4b$ at $m_X = 400\,\mathrm{GeV}$ and non-resonant $\mathrm{ZH} \to 4b$.
\Cref{fig:SR_efficiency_others} shows that the SR also captures a substantial proportion of signal events for these signals once the signal ratio is large enough.
Among the ratios we examined, this occurs from $\epsilon = 0.0075$ onwards for the resonant $\mathrm{HH} \to 4b$ signal and from $\epsilon = 0.03$ onwards for the non-resonant $\mathrm{ZH} \to 4b$ signal, and we display each benchmark at its respective ratio.
These are also the smallest ratios at which the corresponding hypothesis tests attain substantial power, as reported below.
This suggests that our approach to defining the SR can adapt to varying kinematic properties inherent in plausible signal processes, thus preserving the signal concentration characteristics observed in the non-resonant $\mathrm{HH} \to 4b$ scenario.

\begin{figure}
  \centering
  \includegraphics[width=0.7\textwidth]{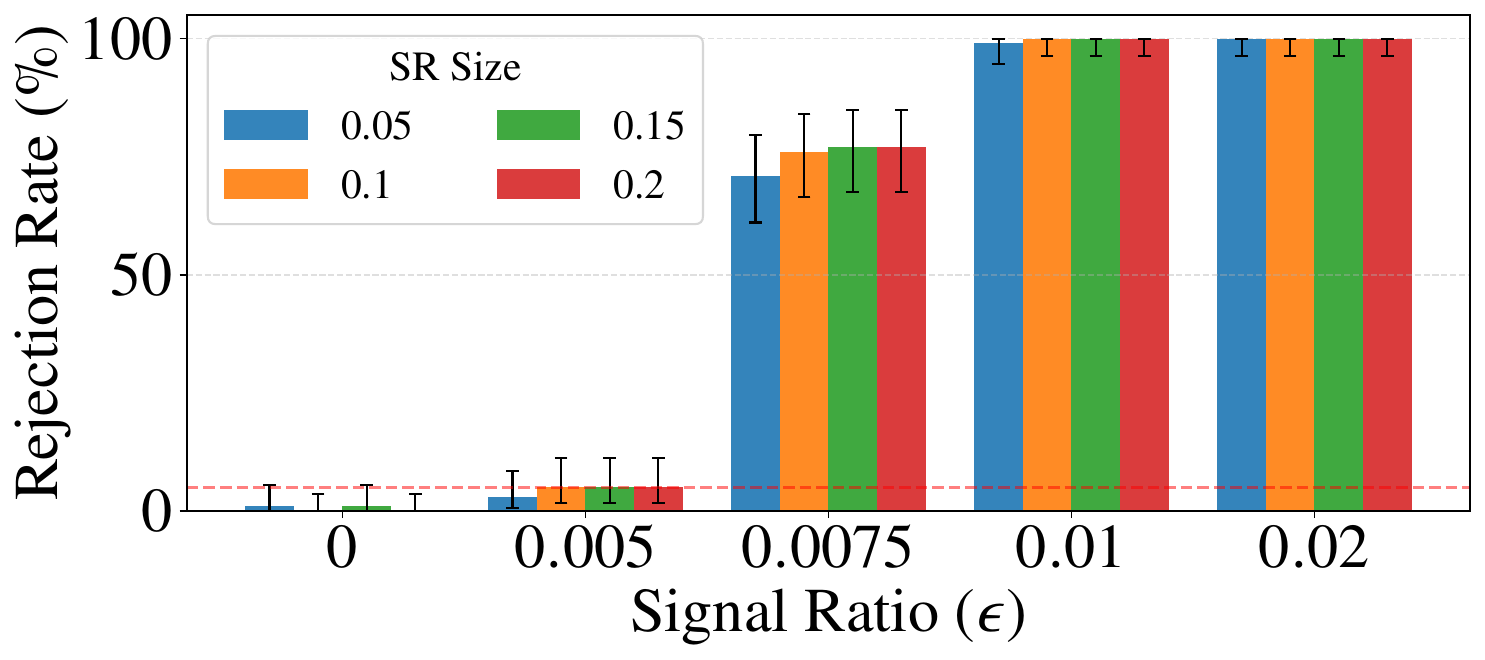}
  \caption{Percentage of experiments rejecting the null hypothesis of \eqref{eq:sr_ratio_test} with resonant $X \to \mathrm{HH} \to 4b$ signal at $400\,\mathrm{GeV}$ for the hypothesis test at significance level $\alpha = 0.05$ with noise scale $\eta = 2.0$.
    Error bars are Clopper--Pearson $95\%$ intervals based on the $100$ repetitions. The power is significant when the signal ratio $\epsilon$ is greater than or equal to $0.0075$.}
    \label{fig:hypothesis_testing_400}
\end{figure}

\begin{figure}
  \centering
  \includegraphics[width=0.7\textwidth]{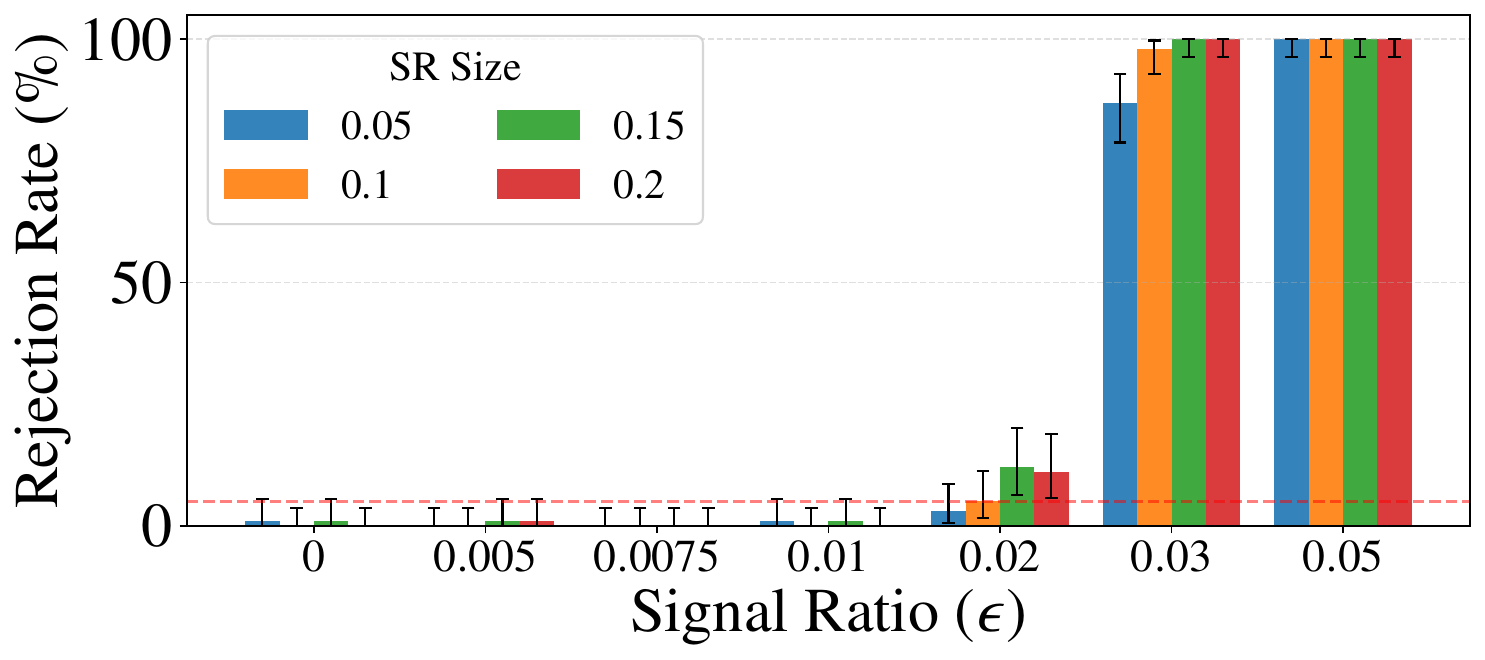}
  \caption{Percentage of experiments rejecting the null hypothesis of \eqref{eq:sr_ratio_test} with non-resonant $\mathrm{ZH} \to 4b$ signal for the hypothesis test at significance level $\alpha = 0.05$ with noise scale $\eta = 2.0$.
    Error bars are Clopper--Pearson $95\%$ intervals based on the $100$ repetitions. The power rises steeply between $\epsilon = 0.02$ and $\epsilon = 0.03$, ranging from $0.87$ to $1.00$ at $\epsilon = 0.03$ and reaching $1.00$ at $\epsilon = 0.05$.}
    \label{fig:hypothesis_testing_zh4b}
\end{figure}

We extend the hypothesis-testing analysis to these two signal processes, using $s_{\text{CR}} = 1 - s_{\text{SR}}$ with $s_{\text{SR}}$ varying across $\{0.05, 0.1, 0.15, 0.2\}$, significance level $\alpha = 0.05$, $B = 1000$ bootstrap replicates, and $100$ repeated experiments drawn as described in \Cref{ss:res_concentration_of_signal_events_in_the_SR}.
When $\epsilon = 0$, no signal is injected, so the null experiment is common to all signal processes. 
The null-rejection results reported in \Cref{ss:res_hypothesis_testing} therefore apply to these experiments as well.
The results for $X \to \mathrm{HH} \to 4b$ at $m_X = 400\,\mathrm{GeV}$ are presented in \Cref{fig:hypothesis_testing_400}.
We observe significant statistical power when the signal ratio $\epsilon \geq 0.0075$, with empirical rejection rates of at least $0.99$ across all considered SR sizes when $\epsilon \geq 0.01$. 
The corresponding performance for the non-resonant $\mathrm{ZH} \to 4b$ signal is shown in \Cref{fig:hypothesis_testing_zh4b}, where power rises steeply between $\epsilon = 0.02$ and $\epsilon = 0.03$, ranging from $0.87$ to $1.00$ at $\epsilon = 0.03$ and reaching $1.00$ at $\epsilon = 0.05$.
Overall, the results for both signal types demonstrate that the proposed model-agnostic framework generalizes robustly across different signal processes without requiring specific knowledge of their detailed kinematic properties.

\section{Discussion}
\label{s:discussion}

In this work, we proposed a model-agnostic framework for discovering new physics in high-energy collision data without strict prior knowledge of either the signal or background.
Our method uses imperfect background events, exemplified by $3b$ events, to define the SR without relying on specific knowledge of the characteristics of the signal.
We observe that an ensemble maximum of classifiers trained to distinguish the misspecified background from the observed \(4b\) events concentrates signal events in the high-density ratio region.
However, a na\"ive definition of the SR by thresholding the estimated density ratio $\widehat{\gamma}$ introduces significant bias in background estimation. We address this issue by introducing a smoothed density-ratio estimate $\hatgammatilde$ constructed by smearing the event representations.
We threshold an ensemble maximum of $\widehat{\gamma}/\hatgammatilde$ to define the SR and CR and estimate the background distribution in the SR by interpolating the density-ratio estimate learned in the CR.
We execute a bootstrap-based KS-type test without introducing a binning hyperparameter in the final test statistic.
Our empirical results show signal concentration in the SR that is competitive with model-aware baselines.
The test in the resulting SR accounts for systematic uncertainty in the fitted classifier through a composite null and exhibits conservative null rejection rates across the settings studied.
Crucially, the method exhibits strong generalization capabilities, extending robustly to both resonant and non-resonant signals across diverse kinematic regimes, hence demonstrating its potential in future model-agnostic searches at the LHC.

An important direction for future research is to establish a principled procedure for selecting the noise scale \(\eta\) used to smear the event representations.
One natural strategy is to perform multiple tests across different values of \(\eta\) and combine their outcomes via a multiple-testing framework, potentially following approaches such as that of \citet{GrossoLetizia2025_multiple}.

%%%%%%%%%%%%%%%%%%%%%%%%%%%%%%%%%%%%%%%%%%%%%%
%% Support information, if any, goes in the %%
%% Acknowledgements section.                %%
%%%%%%%%%%%%%%%%%%%%%%%%%%%%%%%%%%%%%%%%%%%%%%
%\begin{acks}[Acknowledgments]
%The authors thank the anonymous reviewers for their valuable suggestions.
%\end{acks}

%%%%%%%%%%%%%%%%%%%%%%%%%%%%%%%%%%%%%%%%%%%%%%
%% Funding information, if any, goes here.  %%
%%%%%%%%%%%%%%%%%%%%%%%%%%%%%%%%%%%%%%%%%%%%%%
%\begin{funding}
%This work is supported in part by funds from the National Science Foundation
%(NSF: \# 1636933 and \# 1920920).
%\end{funding}

%%%%%%%%%%%%%%%%%%%%%%%%%%%%%%%%%%%%%%%%%%%%%%
%% Supplementary Material, including data   %%
%% sets and code, should be provided in a   %%
%% {supplement} environment with title and  %%
%% short description.                       %%
%%%%%%%%%%%%%%%%%%%%%%%%%%%%%%%%%%%%%%%%%%%%%%
%\begin{supplement}
%\stitle{???}
%\sdescription{???}
%\end{supplement}

%%%%%%%%%%%%%%%%%%%%%%%%%%%%%%%%%%%%%%%%%%%%%%
%%              The Bibliography             %%
%% imsart-nameyear.bst creates the .bbl      %%
%% file that is submitted to AOAS.           %%
%%%%%%%%%%%%%%%%%%%%%%%%%%%%%%%%%%%%%%%%%%%%%%
\bibliographystyle{imsart-nameyear}
\bibliography{reference}

\clearpage

%%%%%%%%%%%%%%%%%%%%%%%%%%%%%%%%%%%%%%%%%%%%%%
%% AOAS asks that appendices be moved to a  %%
%% separate supplementary article. They are %%
%% kept here for the working draft.         %%
%%%%%%%%%%%%%%%%%%%%%%%%%%%%%%%%%%%%%%%%%%%%%%
\appendix

\section{Event Weights}
\label{s:weights}

This section explains how we handle the event weights that accompany the $3b$ and $4b$ samples, a discussion we have deferred from the main text for the sake of clarity.
All event samples considered in this work are modeled to incorporate multiple jets, possibly more than four. To improve the model's accuracy, \citet{Bryant2018_search} applies two forms of reweighting to each modeled event: combinatorial and kinematic reweighting. The former addresses inefficiency of $b$-tagging, and the latter reduces the kinematic discrepancy between the simulated dataset and actual data. Due to the combinatorial perspective of reweighting, the weights are not precisely a function of the observed features $x$. For notational convenience, however, we write $w_{3b}(x)$ and $w_{4b}(x)$ for the weights assigned to a $3b$ or $4b$ event $x$, respectively.

Accounting for the weights modifies several definitions and algorithms from the main text.
We restate the affected ones below, highlighting the changes in red.
Specifically, the density ratio of $4b$ events to $3b$ events is modified to
\begin{equation}\label{eq:density_ratio_weight}
    \gamma(x) \coloneqq \frac{\hP(Y = 1 \mid X = x)}{\hP(Y = 0 \mid X = x)} = 
    \frac{\red{w_{4b}(x)} (b_{4b}(x) + \mu s(x))}{\red{w_{3b}(x)} b_{3b}(x)},
\end{equation}
which incorporates the weights.
This suggests that we should consider the weights in the process of learning the density ratio to align with its definition.
Concretely, every classifier trained in \Cref{alg:learn_density_ratio_and_representations} is fit by minimizing a \emph{weighted} cross-entropy loss, in which each event contributes proportionally to its weight,
\begin{equation}\label{eq:weighted_cross_entropy}
    \cL(\widehat{p}) = -\sum_{i \,:\, Y^i = 1} \red{w_{4b}(X^i)} \log \widehat{p}(X^i) - \sum_{i \,:\, Y^i = 0} \red{w_{3b}(X^i)} \log \left(1 - \widehat{p}(X^i)\right),
\end{equation}
where $\widehat{p}$ estimates $\hP(Y = 1 \mid X = x)$ and the density ratio is recovered as $\widehat{\gamma} = \widehat{p} / (1 - \widehat{p})$.
The population minimizer of \eqref{eq:weighted_cross_entropy} is the weighted class probability, whose odds are precisely the weighted density ratio in \eqref{eq:density_ratio_weight}; training with the weighted loss therefore targets the correct quantity, whereas the unweighted loss would estimate the ratio of the unweighted event densities.
The balancing convention $\hP(Y = 0) = \hP(Y = 1) = 1/2$ of \Cref{s:problem_formulation} is likewise imposed on the total weight of each class rather than on the raw event counts.
In the same context, we use ``counts of events'' in a set to refer to the sum of weights of events in the set.
Before the affine correction is applied, the weight-aware cumulative estimators are
\begin{equation}\label{eq:weighted_empirical_cdf_weight}
\begin{aligned}
    \widehat{F}_{3b}(y) &= \frac{\sum\limits_{X_{3b}^i \in \cX_s} (\widehat{\gamma}_{\text{CR}} \red{w_{3b}})(X_{3b}^i) \,\mathbf{1}\{\widehat\psi(X_{3b}^i) \leq y\}}{\sum\limits_{X_{3b}^i \in \cX_s} (\widehat{\gamma}_{\text{CR}} \red{w_{3b}})(X_{3b}^i)},\\
    \widehat{F}_{4b}(y) &= \frac{\sum\limits_{X_{4b}^j \in \cX_s} \red{w_{4b}}(X_{4b}^j) \,\mathbf{1}\{\widehat\psi(X_{4b}^j) \leq y\}}{\sum\limits_{X_{4b}^j \in \cX_s} \red{w_{4b}}(X_{4b}^j)}.
\end{aligned}
\end{equation}

Some simulated signal samples contain negative event weights arising from higher-order corrections \citep{AndersenGutschowMaierPrestel2020_negative}. 
We retain these weights with their signs in signal-weighted summaries and in the hypothesis test. 
Thus, the finite-sample quantity $\widehat{F}_{4b}$ in \eqref{eq:weighted_empirical_cdf_weight} can locally decrease when such a signal is included; it is then a signed Monte Carlo estimator of the physical cumulative distribution rather than a CDF in the literal finite-sample sense. Multiplication of all weights in either class by a positive constant has no effect on the normalized quantities in \eqref{eq:weighted_empirical_cdf_weight}. The signed-weight pooled resampling procedure is described in \Cref{s:continuous_affine_bootstrap}.

For completeness, \Cref{alg:define_regions_weight} gives the weight-aware version of \Cref{alg:define_regions}, with the changes highlighted in red.

\begin{algorithm}
    \caption{Updated version of \Cref{alg:define_regions} reflecting weights}
    \begin{algorithmic}[1]
        \Procedure{DefineRegions}{$\bX_1$, $\widehat{\psi}$, $s_{\text{SR}}$, $s_{\text{CR}}$}
        \State $\bX_{4b} \gets \{x_{4b}\,|\,x_{4b} \in \bX_1 \text{ and } x_{4b} \text{ is a } 4b \text{ event}\}$, \red{$N_{4b} \gets \sum_{x_{4b} \in \bX_{4b}} w_{4b}(x_{4b})$}
        \State \red{$\tau_s \gets$ weighted empirical $(1 - s_{\text{SR}})$-quantile of $\{\widehat{\psi}(x_{4b})\,|\,x_{4b} \in \bX_{4b}\}$ with weights $\{w_{4b}(x_{4b})\}$}
        \State \red{$\tau_c \gets$ weighted empirical $(1 - s_{\text{SR}} - s_{\text{CR}})$-quantile of $\{\widehat{\psi}(x_{4b})\,|\,x_{4b} \in \bX_{4b}\}$ with weights $\{w_{4b}(x_{4b})\}$}
        \State \Return $\cX_s \coloneqq \widehat{\psi}^{-1}([\tau_s, \infty)), \quad \cX_c \coloneqq \widehat{\psi}^{-1}([\tau_c, \tau_s))$
        \EndProcedure
    \end{algorithmic}
    \label{alg:define_regions_weight}
\end{algorithm}

\section{Details of Hypothesis Testing with Systematic Uncertainty}
\label{s:continuous_affine_bootstrap}

For numerical stability, the implementation clips the log-score to $[-10,10]$.
Throughout this appendix, $z$ denotes this clipped log-score:
\begin{equation}
    z(x)=\min\{10,\max\{-10,\log\widehat\psi(x)\}\}.
    \label{eq:clipped_sr_score}
\end{equation}
The cumulative estimators in \eqref{eq:weighted_empirical_cdf} and \eqref{eq:weighted_empirical_cdf_weight} are evaluated using $z$ in place of $\widehat\psi$, and the affine correction in \eqref{eq:affine_correction} uses these clipped log-scores and their weighted mean.
The SR is selected before clipping, so clipping does not change SR membership.
Let $z_{3b}^i=z(X_{3b}^i)$ and $z_{4b}^j=z(X_{4b}^j)$ for the events in the SR, and define the observed $3b$ bounds
\begin{equation}
    L=\min_i z_{3b}^i,\qquad U=\max_i z_{3b}^i.
    \label{eq:affine_score_bounds}
\end{equation}
Write $v^i=(\widehat\gamma_{\mathrm{CR}}w_{3b})(X_{3b}^i)$ and $w_{4b}^j=w_{4b}(X_{4b}^j)$, with nonnegative $v^i$ and positive original totals $W_3=\sum_i v^i$ and $W_4=\sum_j w_{4b}^j$.
With $A=\sum_i v^i z_{3b}^i/W_3$, the centered correction in \eqref{eq:affine_correction} translates to
\begin{equation}
    v^i(\delta)
    =(\widehat{\gamma}_{\mathrm{CR}}^\delta w_{3b})(X_{3b}^i)
    =v^i\bigl[1+\delta(z_{3b}^i-A)\bigr],
    \qquad \sum_i v^i(\delta)=W_3.
    \label{eq:affine_total_preservation}
\end{equation}
The set of values of $\delta$ for which the affine correction factor is nonnegative on $[L,U]$ is
\begin{equation*}
    \Delta=[\delta^-,\delta^+],
    \qquad \delta^-=-\frac{1}{U-A},
    \qquad \delta^+=\frac{1}{A-L}.
\end{equation*}
The observed statistic is therefore
\begin{equation}
    T_0(\delta)=\sup_y\left|
    \frac{\sum_i v^i(\delta)\mathbf1\{z_{3b}^i\leq y\}}{W_3}
    -\frac{\sum_j w_{4b}^j\mathbf1\{z_{4b}^j\leq y\}}{W_4}\right|.
    \label{eq:affine_observed_statistic}
\end{equation}

For each $\delta\in\Delta$, append the observed $4b$ pairs to the corrected $3b$ pairs to form the pooled sequence $(z^k,u^k(\delta))_{k=1}^{N}$ so that
\begin{equation*}
    \{ (z^k,u^k(\delta)) \,|\, 1 \le k \le N \}
    = \{ (z_{3b}^i, v^i(\delta)) \,|\, 1 \le i \le n_3 \} 
    \cup
    \{ (z_{4b}^j, w_{4b}^j) \,|\, 1 \le j \le n_4 \}.
\end{equation*}
Here $n_3$ and $n_4$ are the numbers of observed $3b$ and $4b$ events in the SR, respectively, with $N=n_3+n_4$.
The pooled weights retain the original class totals $W_3$ and $W_4$.
The Poisson rates are
\begin{equation}
    \lambda_3=\frac{W_3}{W_3+W_4},
    \qquad \lambda_4=\frac{W_4}{W_3+W_4}.
    \label{eq:pooled_poisson_rates}
\end{equation}
For replicate $b$, draw independently
\begin{equation}
    C_3^{k,(b)}\sim\operatorname{Poisson}(\lambda_3),
    \qquad C_4^{k,(b)}\sim\operatorname{Poisson}(\lambda_4),
    \quad k=1,\ldots,N.
    \label{eq:pooled_poisson_counts}
\end{equation}
A pooled event $k$ contributes weights $C_3^{k,(b)}u^k(\delta)$ and $C_4^{k,(b)}u^k(\delta)$ to the two resampled samples.
The total-preservation constraint keeps the rates independent of $\delta$, so the same counts $C_3^{k,(b)}$ and $C_4^{k,(b)}$ can be used for every $\delta$.

For computation, merge the retained entries from the two resampled samples in score order.
For merged entry $m$, let $u_{3b}^{m,(b)}(\delta)$ denote its weight if it belongs to the $3b$ resample and zero otherwise, and define $u_{4b}^{m,(b)}(\delta)$ analogously.
The bootstrap statistic is
\begin{equation}
    T_b(\delta)=\max_\ell\left|
    \frac{\sum_{m\leq\ell}u_{3b}^{m,(b)}(\delta)}{\sum_m u_{3b}^{m,(b)}(\delta)}
    -\frac{\sum_{m\leq\ell}u_{4b}^{m,(b)}(\delta)}{\sum_m u_{4b}^{m,(b)}(\delta)}\right|.
    \label{eq:affine_bootstrap_statistic}
\end{equation}
The maximum is evaluated after each merged entry, including intermediate steps at tied scores.

We compute the pointwise $p$-values using \eqref{eq:affine_bootstrap_ks_main} and maximize over $\delta\in\Delta$.
The maximization over the full interval $\Delta$ can be solved exactly without a discretization grid.
For each replicate $b$, we use the explicit piecewise representations of $T_0(\delta)$ and $T_b(\delta)$ to locate their crossings and determine the set $A_b=\{\delta\in\Delta\,|\,T_b(\delta)\geq T_0(\delta)\}$.
The pointwise $p$-value can then be written as
\begin{equation*}
    p(\delta)=\frac{1+\sum_{b=1}^{B}\mathbf1\{\delta\in A_b\}}{B+1}.
\end{equation*}
The composite $p$-value is therefore $p=(1+M)/(B+1)$, where $M$ is the largest number of sets $A_b$ containing a common $\delta$.

\section{Supplementary Results}

\begin{table}[H]
  \caption{Empirical rejection proportions over $100$ repeated experiments for the test of \eqref{eq:sr_ratio_test} with non-resonant $\mathrm{HH} \to 4b$ signal at significance level $\alpha = 0.05$.
  $\eta$ is the noise scale used in learning the smoothed density ratio $\gammatilde$, and $\epsilon$ is the signal ratio.
  $\epsilon = 0$ corresponds to the null case, for which the observed rejection proportions are below the nominal level.}
  \label{tab:hypothesis_testing_noise_scale}
  \centering
  \begin{tabular}{cc|c|c|c|c|c}
      \toprule
      \(\eta\) & $s_{\text{SR}}$ & $\epsilon = 0$ & $\epsilon = 0.005$ & $\epsilon = 0.0075$ & $\epsilon = 0.01$ & $\epsilon = 0.02$ \\
      \midrule
      \multirow{4}{*}{0.5} & $0.05$ & $0.00$ & $0.00$ & $0.02$ & $0.42$ & $0.75$ \\
        & $0.1$ & $0.00$ & $0.00$ & $0.01$ & $0.43$ & $0.83$ \\
        & $0.15$ & $0.00$ & $0.00$ & $0.03$ & $0.46$ & $0.88$ \\
        & $0.2$ & $0.01$ & $0.01$ & $0.04$ & $0.52$ & $0.93$ \\
        \midrule
      \multirow{4}{*}{1.0} & $0.05$ & $0.01$ & $0.01$ & $0.02$ & $0.50$ & $0.98$ \\
        & $0.1$ & $0.00$ & $0.01$ & $0.03$ & $0.67$ & $0.99$ \\
        & $0.15$ & $0.00$ & $0.01$ & $0.04$ & $0.75$ & $1.00$ \\
        & $0.2$ & $0.02$ & $0.00$ & $0.04$ & $0.74$ & $0.99$ \\
        \midrule
      \multirow{4}{*}{2.0} & $0.05$ & $0.01$ & $0.01$ & $0.05$ & $0.62$ & $0.97$ \\
        & $0.1$ & $0.00$ & $0.01$ & $0.11$ & $0.89$ & $0.99$ \\
        & $0.15$ & $0.01$ & $0.02$ & $0.08$ & $0.87$ & $1.00$ \\
        & $0.2$ & $0.00$ & $0.00$ & $0.12$ & $0.90$ & $1.00$ \\
        \midrule
      \multirow{4}{*}{3.0} & $0.05$ & $0.00$ & $0.00$ & $0.11$ & $0.72$ & $1.00$ \\
        & $0.1$ & $0.00$ & $0.02$ & $0.13$ & $0.95$ & $1.00$ \\
        & $0.15$ & $0.00$ & $0.00$ & $0.15$ & $0.97$ & $1.00$ \\
        & $0.2$ & $0.01$ & $0.00$ & $0.16$ & $0.97$ & $1.00$ \\
        \bottomrule
  \end{tabular}
\end{table}

\end{document}